\documentclass[12pt,draftcls,onecolumn,a4paper,journal]{IEEEtran}

\usepackage{textcomp}
\DeclareMathAlphabet\mathbfcal{OMS}{cmsy}{b}{n}

\usepackage{hyperref}

\usepackage[linesnumbered,vlined,ruled]{algorithm2e}
\usepackage{amsmath,amsfonts,amssymb,amsthm}
\theoremstyle{plain}
\newtheorem{thm}{Theorem}

\newtheoremstyle{cited}%
  {3pt}
  {3pt}
{\itshape}
  {}
  {\bfseries}
  {.}
  {.5em}
  {\thmname{#1} \thmnumber{#2} \thmnote{\normalfont#3}}
\theoremstyle{cited}
\newtheorem{citedthm}[thm]{Theorem}
\newtheorem{citedlem}[thm]{Lemma}
\newtheorem{citedcor}[thm]{Corollary}

\newtheorem{citedprop}[thm]{Proposition}

\usepackage{graphicx}
\usepackage{epstopdf}
\usepackage{epsf}
\usepackage{subfigure}

\usepackage{cite}

\usepackage{tabularx}
\usepackage{array}
\usepackage{booktabs}

\usepackage{comment}

\begin{document}

\title{Frame-based Sparse Analysis and Synthesis Signal Representations and Parseval K-SVD}

\author{Wen-Liang Hwang, Ping-Tzan Huang, and Tai-Lang Jong
  \thanks{Wen-Liang Hwang is at Institute of Information Science, Academia 
  Sinica, Taipei 11529, Taiwan.  Ping-Tzan Huang and Tai-Lang Jong are at Department of Electrical Engineering, National Tsing Hua University, Hsinchu, Taiwan.}}

\date{}
\maketitle

\begin{abstract}

Frames are the foundation of the linear operators used in the decomposition and reconstruction of signals, such as the discrete Fourier transform, Gabor, wavelets, and curvelet transforms. The emergence of sparse representation models has shifted of the emphasis in frame theory toward sparse $l_1$-minimization problems. In this paper, we apply frame theory to the sparse representation of signals in which a synthesis dictionary is used for a frame and an analysis dictionary is used for a dual frame. We sought to formulate a novel dual frame design in which the sparse vector obtained through the decomposition of any signal is also the sparse solution representing signals based on a reconstruction frame. Our findings demonstrate that this type of dual frame cannot be constructed for over-complete frames, thereby precluding the use of any linear analysis operator in driving the sparse synthesis coefficient for signal representation.
Nonetheless, the best approximation to the sparse synthesis solution can be derived from the analysis coefficient using the canonical dual frame. In this study, we developed a novel dictionary learning algorithm (called Parseval K-SVD) to learn a tight-frame dictionary. We then leveraged the analysis and synthesis perspectives of signal representation with frames to derive optimization formulations for problems pertaining to image recovery. Our preliminary, results demonstrate that the images recovered using this approach are correlated to the frame bounds of dictionaries, thereby demonstrating the importance of using different dictionaries for different applications.

\end{abstract}
\pagenumbering{arabic}

\section{Introduction}

Signal representation is a fundamental aspect of signal processing, serving as the basis of signal decomposition (analysis), processing, and reconstruction (synthesis) \cite{Opp97}. A signal is first mapped as a vector of coefficients in the transform domain using a decomposition operator (analysis operator). The coefficients are then modified and mapped to the signal space using a reconstruction operator (synthesis operator). 
The fundamental prerequisite in the design of analysis and synthesis operators is the perfect reconstruction of the signal\cite{Strang96,Mallat08}. 
Frame theory refers to the branch of mathematics tasked with the design of linear decomposition and reconstruction operators capable of perfectly reconstructing any signal in a vector space \cite{Duf52,Young80,Daubechies92,Christensen08}.  
A frame comprises of a set of linearly independent vectors spanning a vector space.
Any signal in the vector space can be represented as a linear combination of elements in the frame. Analysis coefficients (also referred to as decomposition coefficients or frame coefficients) are derived by applying the input product between the signal and elements in a dual frame. A dual frame associated with the frame  used for this type of signal representation is not unique. Milestones in the development of frame theory include the construction of the canonical dual frame via the frame operator, the constructions of discrete Fourier transform, Gabor, wavelets, and curvelet transforms \cite{Daubechies92,Daubechies86, Daubechies90,Candes00}, and the exploration of frame redundancy in various signal processing applications\cite{Mallat08,Dan11, Quan15}. In cases where the signal space is a finite-dimensional discrete vector space, the frame and canonical dual frame have a connection to pseudo-inverse, singular value decomposition in matrix linear algebra. 

The emergence of the sparse representation model has shifted the emphasis in frame theory toward sparse $l_1$-minimization problems\cite{Chen01,Candes06F,Donoho06}, as evidenced by the plethora of algorithms aimed at elucidating the model and making practical use of it.
Sufficient conditions for a unique $k$-sparse solution to the sparse model can be derived through the analysis of mutual incoherence \cite{Donoho06,Tropp04}, the null space property \cite{Coh09}, the restricted isometry property (RIP) \cite{Candes05}, and spark\cite{Donoho03}, $(\psi)$, which is defined as the smallest possible number $k$ such that there exists a subset of $k$ columns of $\psi$ that are linearly dependent. Throughout the paper, we assume that
\begin{equation}
\text{spark}(\psi) > 2 k \label{spark}
\end{equation}
to ensure that the solution to the $l_1$-synthesis problem is unique. We also assume that $\psi$ (denoting a frame) and $\phi$ (denoting a dual frame) are both in $\mathbb{R}^{m \times n}$ with $m \ge n$ and the ranks of $\psi$ and $\phi$ are $n$.

In this paper, we introduce a novel method by which to  establish a connection between sparse representation problems and frame theory. If we regard the synthesis dictionary for sparse representation as a frame, then the analysis dictionary can be regarded as its dual frame.
This novel perspective leads to the following conclusions: (1) It is impossible to construct a dual frame for any over-complete frame with the aim of obtaining the minimizer of the $\ell_1$-synthesis-based problem; and, (2) The canonical dual frame is best linear decomposition operator by which to obtain the approximation of the $\ell_1$-minimizer. 
This provides theoretical support for the trend toward the use of non-linear decomposition operators to derive the $\ell_1$-minimizer in a straightforward manner; i.e., without relying on iterative algorithms\cite{Bar17,Bor17}. 

The relationship between sparse representation problems and frame theory shifts our perspective with regard to dictionary learning problems. To the best of our knowledge, no existing dictionary (or frame) learning algorithm has addressed the issue of frame bounds. Frame bounds correspond to the largest and smallest eigenvalues of the frame operator $\psi\psi^\top$ of frame $\psi$. The ratio of the bounds (i.e., the condition number) determines the degree of  redundancy between frame coefficients\cite{Pei97}, which means that it is fundamentally associated with numerical stability and the performance of various signal processing problems. In this paper, we demonstrate that the properties of frame theory can be applied to the development of a learning algorithm to learn a Parseval dictionary from a set of observations. This is a frame whose numerical properties are closest to an orthonormal matrix, as the ratio of the frame bounds is $1$.
Finally, we describe the preliminary application of the resulting learned dictionary to problems associated with the restoration of images (denoising, image compression, and filling-the-missing-pixels). Our objective is to illustrate how the frame bounds of dictionaries affects the performance of optimization problems that leverage the synthesis and analysis perspectives of frame coefficients of images.

The remainder of the paper is organized as follows. Section \ref{secframe} presents a review of studies on frame theory as well as analysis and synthesis sparsity. 
Our main theoretical contributions are presented in Section \ref{secsparse} and Section \ref{seckernel}. Section \ref{learning} outlines our approach to training a Parseval tight frame from observations. Section \ref{Sec:Experimental results} presents our experiment results. Concluding remarks are drawn in Section \ref{conclusion}.


\section{Background and Related Works}\label{secframe}

\subsection{Frames}  

A frame comprises a set of linearly independent vectors in a vector space spanned by the vectors. Frames are the cornerstone of signal processing in the formulation of perfect reconstruction pairs of linear operators used to  decompose signals into the transform domain and reconstruct the original signals from the transform coefficients.


Frame studies designing a set of vectors $\psi_i \in V$ span the vector space  by representing any $f \in V$  as 
\begin{equation}
f  = \sum_{i=1}^m \langle f, \tilde c_i \rangle \psi_i,
\end{equation}
where $\{\tilde c_i\}$ and  $\{\langle f, \tilde c_i \rangle\}$ are called the dual frame and the frame coefficients of $\psi$, respectively \cite{Duf52, Daubechies92}. If frame $\psi$ is over-complete, then there are different choices for $\tilde g_i \neq \tilde c_i$ in which 
\begin{equation}
f  = \sum_{i=1}^m \langle f, \tilde g_i\rangle \psi_i.
\end{equation}
The frame operator $S$ of frame $f_i$ with $i=1, \cdots, m$ is defined as follows:
\begin{equation}
Sf = \psi^\top \psi f = \sum_{i}^m \langle f, \psi_i \rangle \psi_i,
\end{equation}
where $\psi^\top$  maps signal $f$ in $V$ to  coefficients in $\mathbb{C}^m$.
Ensuring that $S$ has an inverse requires that the frame condition is met for any $f \in V$:
\begin{equation}
A \|f\|^2 \leq \langle f, Sf \rangle = \langle f, \psi\psi^\top f\rangle = \langle \psi^\top f, \psi^\top f \rangle \leq B\|f\|^2, \label{framebound}
\end{equation}
where $\infty > B \geq  A > 0$ are frame bounds. If  $A = B$, then the frame is tight;  and if $A = B = 1$, then the frame is Parseval. 



The canonical dual frame is defined as sequence $S^{-1} f_1, \cdots,S^{-1}f_m$, where $S^{-1}$ is the inverse of the frame operator. The canonical dual frame satisfies the frame condition for any $f \in V$ with bounds $A^{-1}$ and $B^{-1}$:
\begin{equation}
B^{-1} \|f\|^2 \leq \langle f, S^{-1}f\rangle \leq A^{-1}\|f\|^2.
\end{equation}
A milestone of the frame theory indicates that the frame coefficients derived from the canonical dual frame is the solution of $\ell_2$-synthesis problem
\begin{eqnarray}
\left\{\begin{array}{ll}
 \min\limits_{u} \|u\|_2  &\\
 f = \sum_{i=1}^m u_i \psi_i  &\text{ for all $f \in V$},\label{framel21} 
\end{array}\right.
 \end{eqnarray}
with $u = [u_1 \cdots u_m]^\top$. This sequence of coefficients has the smallest $l_2$-norm of all frame coefficients of $\psi$ representing any signal $f \in V$.

\subsection{Analysis and Synthesis Sparse Models}

The synthesis-based sparse model comprises a dictionary $\psi$ of size $n \times m$ with $m \ge n$ in which it is assumed that a signal is a linear combination of fewer than $k$ atoms in the dictionary ($x = \psi u$ and $\| u \|_0 \le k$). Signals of interest lie within a union of $k$-subspaces of the space spanned by all atoms in the dictionary. 
The parallel analysis-based sparse model (also called the co-sparse analysis model) is used to design an analysis operator $\phi$ (a $n \times m$ matrix with $m \ge n$) in which the analysis vector $\phi^\top x$ of signal $x$ is sparse and the signal with the sparsest analysis coefficients is recovered. 

The paper by Elad et al. \cite{Elad07} provided deep insight into the use of  analysis and synthesis models as priors for Bayesian methods. They pointed out that the two models are equivalent as long as $\psi$ is square and invertible. They also provided examples showing the dichotomy between the two models in the case of over-complete dictionaries, where $m > n$. Finally, they presented theoretical results indicating that any sparse analysis-based problem poses an equivalent sparse synthesis-based problem, but the reverse does not hold. They also demonstrated that the reformulation of an analysis problem to an identical synthesis problem can lead to an exponentially large dictionary. Nam et al. \cite{Nam11} demonstrated that, for over-complete $\psi$ and $\phi$, there is generally a considerable difference in the union of subspaces provided by synthesis and analysis models. 
If $\psi$ and $\phi$ both have sizes of $n \times m$, then 
the number of atoms of $\psi$ that can synthesize a signal is freely from $0$ to $n-1$ while  that of $\phi$ that obtains zero coefficients of $\phi^\top x$ (co-sparsity) freely and from $0$ to $n-1$. 
The number of zero coefficients of $\phi^\top x$ is inversely proportional to the subspace of $\psi$ in which signal $x = \psi u$ lies on the number of non-zero coefficients in $u$. 
Thus, the algorithm of the co-sparsity analysis model is designed to derive zero coefficients whereas the algorithm of the synthesis-based sparse model is designed to derive non-zero coefficients. 
The above studies demonstrate that analysis and synthesis models can be viewed as complementary.  This, in turn,  suggests that analysis and synthesis operators and the corresponding recovery algorithms could perhaps be designed in pairs.

\section{Sparse Synthesis Coefficients via Dual Frame} \label{secsparse}

The dual frame of $\psi$ that minimizes the $l_2$-norm frame coefficient of any signal is its canonical dual frame.
We sought to determine whether there exists a universal dual frame of $\psi$ that minimizes the $l_1$-norm frame coefficient for any signal. The existence of $l_1$-norm frame coefficients for any $\psi$ is provided by the following proposition.

\begin{citedprop}[\cite{Christensen08}] \label{thm1}
Let $\psi = [\psi_i \in \mathbb{C}^n]_i$ be a frame for finite-dimension vector space $V$. Given $x \in V$, there exists coefficients $d_i \in \mathbb{C}^m$ such that 
\begin{equation}
x = \psi \; d = \sum_{i=1}^m d_i \psi_i,
\end{equation}
and 
\begin{equation}
\|d\|_1 = \inf \{ d_i | x = \sum_{i=1}^m d_i \psi_i\},\label{sm}
\end{equation}
where $d = [d_1, \cdots, d_m]^\top$ is the vector of sparse synthesis coefficients.
\end{citedprop}
If $\psi$ is under-complete or square with $n \ge  m$, then the answer to our question is affirmative and the canonical dual frame is the solution \cite{Elad07}.
Unfortunately, Theorem \ref{cor-1} shows that this is not the case if $\psi$ is over-complete. 

First, we highlight an interesting result of Nam et al.\cite{Nam01,Nam03}: if $\phi$ is a frame with the form of an $n \times m$ matrix and with columns $\phi_i$ in general positions (any set of at most $n$ columns are linearly independent), then the number $k$ of non-zero coefficients of $\phi^\top x$, where $x$ is a signal in $\mathbb{R}^n$, is at least $m-n+1$.  
Since $m-k$ rows in $\phi^\top x$ are zeros and these rows take up no more than $n-1$ dimensions, $k \ge m-n +1$. In the case of $m > n$, the result excludes the case in which $\phi^\top x$ is sparse for any $k$. 
Theorem \ref{cor-1} extends this result by showing that there does not exist a dual frame $\phi$ of over-complete $\psi$ for any pair of $x \in \mathbb{R}^n$ and $u \in \mathbb{R}^m$ with $k \ge m+ n - 1$ that satisfies 
\begin{eqnarray}
\left\{\begin{array}{lll}
x & = &\psi u\; ; \label{thm11}\\
u & = & \phi^\top x.  \label{thm12}
\end{array}\right.
\end{eqnarray}

\begin{citedlem} \label{thm2}
Let  the $n \times m$ matrix $\phi =[ \phi_1 \cdots \phi_m]$ be a frame in $\mathbb{R}^n$ with columns $\phi_i$ in general positions, and $\phi$ is a dual frame of $\psi$ (thus, $\psi\phi^\top = I_{n \times n}$). 
Suppose that $k \ge m-n+1$.
Then, for $ m > n$, there exists no pair of $\phi$ and $\psi$ that meets (\ref{thm11}) for any pair of $k$-sparse vector $u$ and signal $x$. 
\end{citedlem}

\noindent{{\bf{Proof.}}}  Suppose that there is a pair of frame $\psi$ and dual frame $\phi$ satisfying conditions (\ref{thm11}) for all pairs of signal $x$ and corresponding $k$-sparse vector $u$. With no loss of generality,  we suppose that the indices of non-zero coefficients of $u$ are $1, \cdots, k$. Because $x = \psi u$, $x$ is in the subspace spanned by $\psi_1, \cdots, \psi_k$ (denoted as $V^{\psi}_{a}(k)$). Meanwhile, because $u = \phi^\top x$, $x$ is in the subspace spanned by $\phi_1, \cdots, \phi_k$ (denoted as $V_{a}^{\phi}(k)$), as well as in the subspace perpendicular to that spanned by the remaining $m-k$ vectors (denoted as
$V_{b}^{\phi}(m-k)^{\perp}$).  Hence,
\begin{equation}
x \in V_{a}^{\psi}(k) \cap V_{a}^{\phi}(k) \cap V_{b}^{\phi}(m-k)^{\perp}. \label{ybelong}
\end{equation}
Nam et al. \cite{Nam01,Nam03} reported that to have a non-empty intersection of $V_{a}^{\phi}(k)$ and $V_{b}^{\phi}(m-k)^{\perp}$, $k$ must be at least $m-n+1$. Because the set of signals satisfying Equation (\ref{ybelong}) is a vector space of dimension $k$ ($x = \psi u$ and the bottom $m-k$ elements in $u$ are zeros.), the dimension of $V_{b}^{\phi}(m-k)$  is $n-k$.

Let $\phi = [\phi_a\; \phi_b]$ and $\psi  = [\psi_a\; \psi_b]$, where $\phi_a = [\phi_1 \cdots \phi_k]$, $\phi_b = [\phi_{k+1} \cdots \phi_m]$, $\psi_a = [\psi_1 \cdots \psi_k]$, and $\psi_b =[\psi_{k+1} \cdots \psi_m]$. 
Equation (\ref{thm11}) stipulates that  we have
\begin{equation}
u = \phi^\top \psi u.
\end{equation}
Therefore, $\phi_b^\top \psi_a u_1 = 0_{m-k}$ and $\phi_a^\top \psi_a u_1 = u_1$,  where $u_1$ is the first $k$ elements in $u$.  Since $u_1 \in \mathbb{R}^k$ is arbitrary, 
we obtain the following:
\begin{eqnarray}
\phi_b^\top \psi_a & =& 0_{m-k \times k};  \label{null}\\
\phi_a^\top \psi_a &= &I_{k \times k}; \label{one}
\end{eqnarray}
which corresponds to 
$V_{b}^{\phi}(m-k) \perp V_{a}^{\psi}(k)$ and $V_{a}^{\phi}(k) = V_{a}^{\psi}(k)$, respectively. Combining  the results of Equations (\ref{null}) and (\ref{one}) while considering that the dimensions of $V_{b}^{\phi}(m-k)$ and  $V_{a}^{\psi}(k)$ are respectively $n-k$ and $k$, we obtain the following:
\begin{equation}
V_{a}^{\psi}(k)  = V_{a}^{\phi}(k) = V_{b}^{\phi}(m-k)^{\perp}. \label{spaceeq}
\end{equation}

Based on the fact that the dimension spanned by $m-k$ columns in $\phi_b$  is $n-k$, we know that $m-k = n-k + (m-n)$. Because $m-n > 0$, we can deduce that there exists a set of $n-k + 1$ vectors in $\phi_b$ spanning a dimension $n-k$ subspace. This violates the assumption that $\phi_i$ are in general positions. \\
\noindent{{\bf{End of Proof.}}}

\begin{citedthm} \label{cor-1}
Let $\psi$ be an over-complete frame and let $u_1^*(x)$ be the $k$-sparse minimizer of $l_1$-synthesis problem for signal $x$ with 
\begin{eqnarray}
\left\{\begin{array}{ll}
\min\limits_u \|u\|_1  &   \label{reconst0}\\
 x = \psi u;  &  
 \end{array}\right.
\end{eqnarray} for signal $x$. There does not exist a dual frame $\phi_1$ in general positions that yields $u_1^*(x)= \phi^\top_1 x$ for any $x $, regardless of the value of $k$. 
\end{citedthm}
\noindent{{\bf{Proof.}}}
Th results reported by Nam et al.\cite{Nam01,Nam03} forfeit the existence of $\phi$ for $k = \| u_1^* \|_0 \le m-n$. Meanwhile, Lemma \ref{thm2} forfeits the existence of $\phi$ for $k \ge m-n +1$. \\
\noindent{{\bf{End of Proof.}}}

Proposition \ref{thm1} claims the existence of $l_1$-norm minimizer for any $x$. Theorem \ref{cor-1} shows that the minimizer cannot be derived through the decomposition of signals with a universal linear operator, which depends exclusively on $\psi$.

\section{Optimal Proxy to Sparse Synthesis Coefficients} \label{seckernel}

Theorem \ref{cor-1} provides a negative answer concerning the existence of a dual frame $\phi_1$ of over-complete frames as the analysis operator to obtain the $l_1$-minimizers of any signal.
Nevertheless, this section presents an affirmative answer concerning the existence of a universal dual frame for the following approximation problem:
\begin{eqnarray}
\left\{\begin{array}{l}
\min\limits_{\phi} \| \phi^\top x - u_1^*(x)\|_2  \label{proxy}\\
\phi \text{ is a dual frame of $\psi$}.
\end{array} \right.
\end{eqnarray}
The following proposition shows that the canonical dual frame, $\phi_2$, is the solution to (\ref{proxy}) resulting in frame coefficients closest to the $l_1$-minimizer of any signal and can be derived by solving the following $\ell_2$-analytical problem:
\begin{eqnarray} \label{analysisl2}
\left\{\begin{array}{l}
 \min\limits_{\phi} \|\phi^\top x\|_2  \\
 \text{$\phi$ is a dual frame of $\psi$}. 
\end{array}\right.
\end{eqnarray}

\begin{citedthm} \label{prop3}
(i) The canonical dual frame yields the optimal proxy of $u_1^*(x)$, the solution of $l_1$-synthesis problem (\ref{reconst0}), for any signal $x$. (ii)  $\phi_2$ is also the minimizer of the $l_2$-analysis problem (\ref{analysisl2}).
\end{citedthm}

\noindent{{\bf{Proof.}}}

(i) If $\psi$ is an under-complete or square frame, then $\phi_2^\top x = u_1^*(x)$ is applicalbe to any $x$ \cite{Elad07}. 

The canonical dual frame, $\phi_2$, of the over-complete frame $\psi$ is $(\psi \psi^\top)^{-1} \psi$.
Since $\phi_2^\top \psi = \psi^\top (\psi \psi^\top)^{-1} \psi = \psi^\top \phi_2$, the kernel $\phi_2^\top \psi$ is the rank $n$ orthogonal projection\footnote{$P$ is an orthogonal projection if and only if $P^2 = P$ and $P^\top =P$.} from $\mathbb{R}^m$ to $\mathbb{R}^m$, due to the fact that
\begin{eqnarray}
\left\{\begin{array}{ll}
(\phi_2^\top \psi)^2 = \phi_2^\top (\psi \phi_2^\top) \psi = \phi_2^\top \psi ;& \\
(\phi_2^\top \psi)^\top = \psi^\top \phi_2 = \phi_2^\top \psi. &  \label{ortho2}
 \end{array}\right.
\end{eqnarray}
Consequently, the optimal value of problem (\ref{proxy}) for any signal $x$ is \begin{equation}
\| \phi_2^\top x - u_1^*(x)\|_2 = \|\phi_2^\top \psi u_1^* - u_1^*(x)\|_2 = 
\|(\mathcal I_{m\times m} - \phi_2^\top \psi) u_1^*(x)\|_2.
\end{equation} 

If $\phi$ differs from $\phi_2$, then kernel $\phi^\top \psi$ is a projection operator but is not necessarily an orthogonal projection as $(\phi^\top \psi)^2 = \phi^\top (\psi \phi^\top) \psi =\psi^\top \psi$. Thus, the canonical dual frame yields the optimal proxy of $u_1^*(x)$.

\medskip
(ii) The case where $\psi$ is a square or under-complete frame:  Equation $x = \psi u_2^*$, with $u_2^*$ being the minimizer of $\ell_2$-synthesis problem (\ref{framel21}), implies that $(\psi \psi^\top)^{-1} \psi^\top x = u_2^*$.  Since the canonical dual frame $\phi_2^\top$ is $ (\psi \psi^\top)^{-1} \psi^\top$, we have $\phi_2^\top x = u_2^*$. 

In the case where $\psi$ is an over-complete frame: 
$u_2^*$ is also the solution of 
\begin{eqnarray}
\left\{\begin{array}{ll} 
 \min_u \frac{1}{2} \|u\|^2_2 & \\
 x = \psi u. &
\end{array}\right.
\end{eqnarray}
Taking partial derivatives with respect to $\lambda$ and $u$ on the Lagrangian function gives
\begin{equation}
L(u, \lambda) = \frac{1}{2} \|u\|^2_2 + \lambda^\top (\psi u  - x)
\end{equation}
and then setting the results to zero, we obtain solution $(u_2^*, \lambda^*)$ that satisfy,
\begin{equation}
\psi u_2^*  =   x \text{  and  }
u_2^*   =  \psi^\top \lambda^*;
\end{equation}
respectively. It follows that 
$x = \psi \psi^\top \lambda^*.$
$(\psi \psi^\top)^{-1} x  = \lambda^*$ and $\phi_2^\top x\ = u_2^*$ can be deduced from the fact that $\psi$ is an over-complete frame and $\phi_2^\top = \psi^\top (\psi \psi^\top)^{-1}.$

\noindent{{\bf{End of Proof.}}}

\noindent Thus, $\phi_2^\top x$ is referred to as the optimal proxy of $u_1^*(x)$. 
The following corollary is summarized from Theorems \ref{cor-1} and \ref{prop3}.

\begin{figure}
\begin{center}
\includegraphics[width=0.5\textwidth]{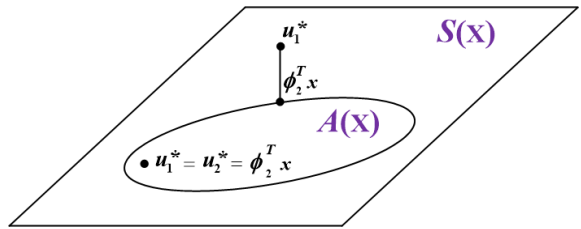}
\caption{Signal $x$ and frame $\psi$ are fixed. $\mathcal S(x)$ is a linear sub-space. $\mathcal A(x)$ is a convex set due to the fact that if $\phi_a$ and $\phi_b$ are dual frames of $\psi$, so does $\alpha \phi_a + (1-\alpha) \phi_b$ for any $\alpha \in [0,1]$. If $u_1^*$ does not belong to $\mathcal A(x)$, then $u_2^*$ is its best approximation in $\mathcal A(x)$; otherwise, $u_1^* = u_2^*$, obtainable by a linear operation.}
\label{fig:xxx}
\end{center}
\end{figure}

\begin{figure}[tp]
\centering
  \includegraphics[width=0.6\textwidth]{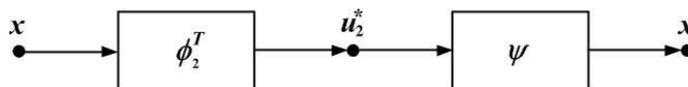}
  \caption{Frame $\psi$ is used as the reconstruction (synthesis) operator. The decomposition (analysis) operator is the transpose of canonical dual frame $\phi_2$. The optimal solution of $\ell_2$-sythesis poblem, $u_2^*$, can be obtained analytically with linear operator $\phi_2$.} \label{signalflow}
\end{figure}

\begin{citedcor}
(i) Let $\psi$ be an over-complete frame; $\phi$ be its dual frame; and let $x$ be a signal. Let ${\cal S}(x) = \{u | x = \psi u\}$ be the synthesis coefficients of $x$ and let ${\cal A}(x) = \{ \phi^\top x | \text{$\phi$ is a dual frame of $\psi$} \}$ be the frame coefficients of the signal. Then, ${\cal A}(x) \subseteq {\cal S}(x)$, $u_1^*(x) \in S(x)$. The orthogonal projection of $u_1^*(x)$ to ${\cal A}(x)$ obtains the frame coefficient $u_2^*(x)$, which is the solution of $\ell_2$-synthesis problem (\ref{framel21}) (see Figure \ref{fig:xxx}). \\
(ii) $u_2^*$ can be obtained linearly by applying $\phi_2^\top$ to $x$ (see Figure \ref{signalflow}).
\end{citedcor}

The fact that $u_1^*$ cannot be obtained with a linear operator implies that some non-linear method must be adopted in order to obtain $u_1^*$ analytically. This conclusion provides theoretical support for the recent trend of adopting non-linear operators to derive the solutions to sparse representation and compressed sensing problems.

\section{Learning Parseval Frames for Sparse Representation} \label{learning}


Aharon et al. \cite{Aharon06} addressed the problem of learning a synthesis dictionary to sparsely represent a set of observations. 
The K-SVD algorithm is perhaps the most popular algorithm of that kind. Let $Y= [y_1, \cdots, y_N]$ be a $n \times N$ matrix, where each column is an observation. The K-SVD algorithm optimizes 
\begin{eqnarray}
\left\{\begin{array}{ll} \label{k-svd0}
\min_{X, \psi} \|Y - \psi X\|_F^2&\\
X = [x_1, \cdots, x_N] & \\
\| x_i \|_0 \le k \text{ for $i= 1, \cdots N$}. &
\end{array}\right.
\end{eqnarray}
Rubinstein et al. \cite{Rub13} addressed a parallel problem, wherein they concentrated on learning an analysis dictionary to produce a sparse outcome from a set of observations. The synthesis and analysis dictionaries are learned using a similar approach; however, they are mutually exclusive. 

Dictionary learning problems can be brought within the context of frame theory by controlling the frame bounds of the learned dictionary. 
Frame bounds $A$ and $B$ in (\ref{framebound}) are related to the correlation between basis elements in frames. Higher $\frac{B}{A}$ values increase the likelihood of correlation between basis elements in frames. 
If $\frac{B}{A}$ is far from $1$, then uniform distortion in a signal yields non-uniform, direction-dependent distortion in its frame coefficients.
The K-SVD algorithm is able to learn a frame $\psi_{ksvd}$; however, it has no control over the bounds, due to the fact that it imposes no constraints on frame bounds (the largest and smallest singular values of $\psi_{ksvd}\psi_{ksvd}^\top$). 
Parseval tight frames with frame bounds of $1$ have been widely used in signal processing\cite{Ron95,Shen06,Don16}. 
Thus, we re-formulated the K-SVD optimization problem in our development of a learning algorithm to learn the optimal Parseval tight frame as well as the sparse coefficients from a set of observations. Note that $\psi$ is a Parseval tight frame if and only if $|\langle\psi^\top x, \psi^\top x\rangle | ^2 = \|x\|^2$. Thus, $\psi \psi^\top = \mathcal I_{n \times n}$.

\subsection{Design of Parseval Frames}

By imposing a Parseval frame $\psi$ using a K-SVD-like approach, we obtain the following:
\begin{eqnarray}
\left\{\begin{array}{ll} \label{k-svd}
\min\limits_{X, \psi} \|Y - \psi X\|_F^2&\\
X = [x_1, \cdots, x_N]   & \\
\| x_i \|_0 \le k \text{ for $i= 1, \cdots, N$} & \\
\psi \text{ is a Parseval tight frame and } \psi \psi^\top = \mathcal I_{n \times n}. & 
\end{array}\right.
\end{eqnarray}

This problem is difficult to solve because applying the augmented Lagrangian approach would introduce fourth-order polynomial terms on $\psi$, due to the tight frame constraint $ \psi \psi^\top = \mathcal I_{n \times n}$. Thus, we take advantage of the fact that the canonical dual frame of a Parseval tight frame is itself 
$(\phi_2 = [ (\psi \psi^\top)^{-1}]^\top \psi  = \psi)$
and formulate (\ref{k-svd}) as follows:
\begin{eqnarray}
\left\{\begin{array}{ll} \label{analysis0k-svd}
\min\limits_{X, \phi, \psi} \|\phi^\top Y - \phi^\top \psi X\|_F^2&\\
X = [x_1, \cdots, x_N]     & \\
\| x_i \|_0 \le k \text{ for $i= 1, \cdots, N$} & \\
\phi = \psi  & \\
\psi \phi^\top = \mathcal I_{n \times n}. & 
\end{array}\right.
\end{eqnarray}
In the following, we present the Parseval-frame learning algorithm, also referred to as the Parseval K-SVD algorithm,  which trains dictionaries as well as sparse coefficients based on the formulation of (\ref{analysis0k-svd}).

\subsection{Parseval K-SVD}

The learning algorithm is designed through the alternating optimization of $(\phi, \psi)$ and $X$.  Parameter $\rho_1$ is introduced and the objective function of the problem is altered to make it the weighted sum of (\ref{k-svd}) and (\ref{analysis0k-svd}), with the aim of enhancing numerical stability in the optimization process, as follows:
\begin{equation} \label{objective0}
 \|\phi^\top Y - \phi^\top \psi X\|_F^2 + \rho_1 \|Y - \psi X\|_F^2.
 \end{equation}
The alternation does not change the minimizers, due to the fact that problems (\ref{k-svd}) and (\ref{analysis0k-svd}) are equivalent\footnote{$\phi^\top Y = \phi^\top \psi X$ if and only if $Y = \psi X$ because of $\psi \phi^\top = \mathcal I$.}. However, as we shown below, $\rho_1$ makes the updated $X$ numerically stable, due to the fact that the nominator of (\ref{equ:12}) is a non-zero value. 

The augmented Lagrangian function of problem (\ref{analysis0k-svd}) is
\begin{eqnarray}
L_{\rho_2, \rho3}(X, \phi, \psi; \lambda_2, \lambda_3) & = 
&\rho_1\left \| Y- \psi X \right \|_F^2 +\left \|  \nonumber \phi^\top Y-\phi^\top \psi X\right \|_F^2 \\ 
 &+&Trace (\lambda_2^\top(\psi \phi^\top - \mathcal I_{n \times n}))+\frac{\rho_2}{2}\left\|\psi \phi^\top-\mathcal I_{n \times n}\right\|_F^2\\
 &+&Trace (\lambda_3^\top(\psi - \phi))+\frac{\rho_3}{2}\left\|\psi-\phi\right\|_F^2,\nonumber
\end{eqnarray}
where $\rho_2$ and $\rho_3$ are parameters, and $\lambda_2$ and $\lambda_3$ are matrices of Lagrangian multipliers of sizes $n \times n$ and $n \times m$, respectively. We solve the following optimization problem by minimizing the primal variables and maximizing the dual variables as follows:
\begin{eqnarray}
\left\{\begin{array}{ll} \label{AL}
\max\limits_{\lambda_2, \lambda_3} \min\limits_{X, \psi, \phi} L_{\rho2, \rho_3} (X, \phi, \psi; \lambda_2, \lambda_3), \\
\| x_i \|_0 \le k \text{ for $i= 1, \cdots N$}.
\end{array}\right.
\end{eqnarray}

The learning algorithm is derived based on the primal-dual approach, in which the optimal Parseval tight frame is learned from image blocks by the alternating direction method of multiplier (ADMM) method \cite{Gabay76, Boyd11}.

We adopted the alternating direction method of multiplier (ADMM) due to the robustness of the updates of dual variables and that fact that this method supports decomposition of primal variable updates. The updates of $\phi$ and $\psi$ are different from that of $X$, due to the fact that the latter has sparse constraints on its columns.  The updates of $\phi$ and $\psi$ are 
\begin{eqnarray}
\phi^{k+1}&   = & \arg\min_{\phi} L_{\rho_2, \rho_3}(X^k, \phi, \psi^k; \lambda_2^k, \lambda_3^k); \label{phiupdate}\\
\psi^{k+1}&   = & \arg\min_{\psi} L_{\rho_2, \rho_3}(X^k, \phi^{k+1}, \psi; \lambda_2^k, \lambda_3^k); \label{psiupdate}
\end{eqnarray}
and the update of $X$ is 
\begin{eqnarray}
\left\{\begin{array}{ll} \label{AL0}
X^{k+1}  =  \arg\min\limits_{X}  L_{\rho_2, \rho_3}(X, \phi^{k+1}, \psi^{k+1}; \lambda_2^k, \lambda_3^k) \label{Xupdate} \\
\| x_i^{k+1} \|_0 \le  k \text{ for $i= 1, \cdots N$}.
\end{array}\right.
\end{eqnarray}
The dual variables $\lambda_2$ and $\lambda_3$ are updated using the standard gradient ascent method of ADMM ($\lambda_2^0$ and $\lambda_3^0$ are initialized using zero matrices):
\begin{align}\label{equ:9}
\begin{split}
\lambda_2^{k+1} & \leftarrow \lambda_2^k+\rho_2(\psi^{k+1}(\phi^{k+1})^\top-\mathcal I_{n \times n}); \\
\lambda_3^{k+1} &  \leftarrow \lambda_3^k+\rho_3(\psi^{k+1}-\phi^{k+1}).
\end{split}
\end{align}
The complexity of updating $\lambda_2$ and $\lambda_3$ are $\mathcal O(n^2)$ and $\mathcal O(nm)$, respectively.  
In the following, we detail the procedures involved in minimizing the primal variables in (\ref{phiupdate}), (\ref{psiupdate}), and (\ref{Xupdate}), wherein we omit the superscript indices on all variables in order to simply the notation. 
\subsubsection{Update of $\phi$ and $\psi$}

Taking the partial derivative of the augmented Lagrangian $L_{\rho_2, \rho_3}(X, \phi, \psi; \lambda_2, \lambda_3)$ with respect to $\phi$ and setting the result to zero, we obtain the following:
\begin{align}
&(2YY^\top-2YX^\top \psi^\top+2 \psi XX^\top \psi^\top-2 \psi XY^\top)\phi+\phi(\rho_2\psi^\top\psi+\rho_3\mathcal I) \nonumber \notag\\
&=-\lambda_2^\top\psi+\rho_2\psi+\lambda_3+\rho_3 \psi.
\label{equ:7}
\end{align}
Similarly, taking the partial derivative of the augmented Lagrangian with respect to $\psi$ and setting the result to zero, we obtain
\begin{align}
&2\phi\phi^\top \psi+\psi(2{\rho_1} XX^\top+ {\rho_2} \phi^\top\phi+ \rho_3\mathcal I)(XX^\top)^{-1} \nonumber \\
&=(2\rho_1YX^\top-\lambda_2\phi + \rho_2\phi - \lambda_3+\rho_3\phi+2\phi\phi^\top YX^\top)(XX^\top)^{-1}.
\label{equ:8}
\end{align}  
Both (\ref{equ:7}) and (\ref{equ:8}) are derived from Sylvester matrix equations:
\begin{equation}
A_1\beta + \beta B_1 =C_1,
\label{equ:18}
\end{equation}
the solutions of which can be derived by solving linear systems using the least square method obtained by taking the $vec$ operator on both sides of (\ref{equ:18}):
\begin{equation}
(\mathcal I_{m \times m} \otimes A_1 + B_1^\top \otimes \mathcal I_{n \times n})vec (\beta) = vec(C_1),
\end{equation}
where $\otimes$ is the Kronecker product, $A_1$, $B_1$ and $C_1$ are known matrices, and $\beta$ is the unknown term.  The $A_1$, $B_1$, $C_1$ and $\beta$ corresponding to (\ref{equ:7}) are $2YY^\top-2YX^\top \psi^\top+2 \psi XX^\top \psi^\top-2 \psi XY^\top$, $\rho_2\psi^\top\psi+\rho_3\mathcal I$, $-\lambda_2^\top\psi+\rho_2\psi+\lambda_3+\rho_3 \psi$, and $\phi$, respectively.  Likewise, $A_1$, $B_1$, $C_1$, and $\beta$ of (\ref{equ:8}) are respectively $2\phi\phi^\top$, $(2{\rho_1} XX^\top+ {\rho_2} \phi^\top \phi+ \rho_3I)(XX^\top)^{-1}$, $(2\rho_1YX^\top -\lambda_2\phi + \rho_2\phi - \lambda_3+\rho_3\phi+2\phi\phi^\top YX^\top)(XX^\top)^{-1}$, and $\psi$. The numbers of constraints and variables for both systems are equal to $mn$. The complexity of updating $\phi$ and $\psi$ by solving systems of linear equations is thus no more than $\mathcal O(m^3n^3)$.

\subsubsection{Update of $X$}

The objective related to the update of $X$ is (\ref{objective0}), the same at that to update $\phi$ and $\psi$.
Following the approaches in \cite{Aharon06,Pen15}, we iteratively and exclusively update the non-zero coefficients in one row of $X$. This ensures that the number of zero coefficients is increased as the number of iterations is increased, because once a coefficient becomes zero, it remains at zero thereafter. 

First, for each row of $X$, we generate a row vector that contains only the non-zero entries in the row.
Let $\psi_k$, $(\phi^\top \psi)_k$ denote the $k$-th columns in $\psi$ and $\phi^\top\psi$, respectively; and let the $1 \times N$ vector $r_k^\top$ denote the $k$-th row in $X$. Furthermore, let $p(i)$ be the $i$-th non-zero index in $r_k^\top$ and let $\left\|r_k^\top \right\|_0$ be the number of non-zero coefficients. If we let $G_k$ be an $N \times\left\|r_k^\top \right\|_0$ matrix in which $(p(i),i)$ is set at one and the other entries are set at zero, then $r_k^\top G_k$ is a 1 $\times \left\|r_k^\top \right\|_0$ vector with non-zero entries in row $r_k^\top$ of $X$. 
For example, if $r_k^\top=[0\; 0 \;0 \;2\; 0\; 0\; 1]$, then $r_k^\top G_k$ is $[2\; 1]$. 

Next, we consider the update of $r_k^\top G_k$. Let $B=\phi^\top Y$; let $\hat r_k^\top = r_k^\top G_k$; and let $E_k=Y-\sum_{j\neq k}\psi_jr_j^\top$, $F_k=B-\sum_{j\neq k}\psi_jr_j^\top$, $\tilde{E_k}=E_kG_k$, and $\tilde{F_k}=F_kG$.  $E_k$, $F_k$, and $\tilde{E_k}$ are all known matrices. Then, we obtain the following
\begin{align*}
&\rho_1\left\|YG_k-\psi XG_k\right\|_F^2+\left\|\phi^\top YG_k-\phi^\top \psi XG_k\right\|_F^2 \\
&=\rho_1\left\|YG_k-\psi XG_k\right\|_F^2+\left\|BG_k-\phi^\top \psi XG_k\right\|_F^2 \\
&=\rho_1\left\|(Y-\sum_{j\neq k}\psi_j r_j^\top)G_k-\psi_k r_k^\top G_k\right\|_F^2 + \left\|(B-\sum_{j\neq k}\psi_j r_j^\top)G_k-\phi^\top\psi_k r_k^\top G_k\right\|_F^2\\
&=\rho_1 \left\|E_kG_k-\psi_k r_k^\top G_k\right\|_F^2 + \left\|F_kG_k-\phi^\top \psi_k r_k^\top G_k\right\|_F^2\\
&=\rho_1 \left\|\tilde{E_k}-\psi_k \hat r_k^\top \right\|_F^2 +\left\|\tilde{F_k}-\phi^\top \psi_k\hat r_k^\top \right\|_F^2.
\end{align*}
The non-zero coefficients in the $k$-th row of $X$ can be updated by solving  
\begin{equation*}
\min\limits_{\hat r_k^\top} \rho_1\left\|\tilde{E_k}-\psi_k \hat r_k^\top \right\|_F^2 +\left\|\tilde{F_k}-\phi^\top \psi_k\hat r_k^\top \right\|_F^2.
\end{equation*}
The above has the following closed-form solution:
\begin{align}
 \hat r_k^\top =\frac{\rho_1 \psi_k^\top \tilde{E_k}+\psi_k^\top\phi \tilde{F_k}}{\rho_1\left\|\psi_k\right\|^2+\left\|\phi^\top\psi_k\right\|^2}.
\label{equ:12}
\end{align}
Thus, the complexity of one update of non-zero entries in a row of $X$ takes $\mathcal O(n^2)$.
This update of non-zero entries in a row is processed from row $1$ to row $m$ and repeats $j$ times in one update of $X$. A full description of the proposed learning algorithm is given in Table \ref{table1}.

{
\vspace{0.5cm}

\begin{table}[!ht]
\caption{Parseval K-SVD} \label{table1}
\begin{tabular}{@{}p{14cm}@{}}
\hline
Parseval K-SVD Algorithm \\
\hline
Deriving initial frame $\psi$ and coefficients $X$ from observation $Y$ using the K-SVD with DCT as the initial dictionary of K-SVD.\\
\hline
{\bf Input:}\\
(i) Set initial $\psi$ (of size $n \times m$) and $X$ (of size $m \times N$).\\
(ii) Set the values for ${\rho_1, \rho_2, \rho_3,}$ and the maximum number of iterations. \\
(iii) The initial matrices of Lagrangian multipliers $\lambda_2$ and $\lambda_3$ are set to zero.\\
(iv) The default iteration number $j = 20$ for one update of $X$ . \\
For i = 1 to max number of iterations: \\

\quad (v) Update $\phi$ by solving (\ref{equ:7}) followed by updating $\psi$ using (\ref{equ:8}).\\
\quad (vi) Update Lagrangian matrices $\lambda_2$ and $\lambda_3$ using (\ref{equ:9}).\\
\quad (vii) Update $X$ one row at a time from the first to the last row. This process is repeated $j$ times.\\ 
end i \\
{\bf Output:}\\
$\psi$, $\phi$, and $X$\\   
\hline
\end{tabular} 
\end{table}

}
The complexity associated with one update of $\phi$ and $\psi$ of Step (v) involves solving systems of $mn$ linear equations, each of which takes at most $\mathcal O(m^3 n^3)$. Meanwhile, one update of non-zero entries in $X$ in Step (vii) takes $\mathcal O(jn^2 m)$, where $j$ (default is set at $20$) is the number of iterations associated with one update of $X$. The complexity of updating $\lambda_2$ and $\lambda_3$ in Step (vi) are $\mathcal O(n^2)$ and $\mathcal O(n^2m)$, respectively. Thus, one update of Parseval K-SVD (Steps (v), (vi) and (vii)) is $\mathcal O(m^3 n^3)$. Let $M$ (typically, $M=200$) be the number of iterations  required for the learning algorithm to converge. The complexity of learning the Parseval dictionary is thus $\mathcal O(m^3 n^3 M) + $ the complexity of learning K-SVD dictionary.  However, the Parseval K-SVD algorithm needs to be completed only once, after which the learned dictionary can be used for a host of sparse recovery problems.

Finally, we address the question of whether the Parseval K-SVD algorithm converges. First, let us assume that $X$ is fixed. Each update of either $\phi$ or $\psi$ would decrease the objective value of (\ref{objective0}) or result in no change. Then, with fixed $\phi$ and $\psi$, updating non-zero entries in each row of $X$ would reduce this objective value as well as the number of zero elements in the row, or result in no change.  
Executing a series of such steps would ensure a monotonically non-increasing of the objective value, thereby guaranteeing convergence to a local minimum (\ref{objective0}).

\section{Experiment Results}\label{Sec:Experimental results}

We first applied the Parseval K-SVD algorithm to blocks of natural images in order to determine whether the learned dictionary is capable of recovering the desired results. We then conducted several experiments using natural image data, in an attempt to demonstrate the applicability of the synthesis and analysis views of image representation when used in conjunction with the learned dictionaries (K-SVD and Parseval K-SVD) for the restoration of images.

\subsection{Learning Parseval Dictionaries and Reconstruction of Original Images}

Our goal is to derive a Parseval dictionary capable of recovering original images using linear operators, based on the tenets of frame theory.
We consider a number of implementation issues. The matrix of observations $Y$ is of size $n \times N$, set at $64 \times 4096$. The initial dictionary $\psi$ and coefficients $X$ in Table \ref{table1}(i) are derived using the K-SVD method.  The dictionary $\psi$ is of size $n \times m$, set at $64 \times 256$; and the coefficient $X$ of size $m \times N$ is set at $256 \times 4096$. The training data consists of $4,096$ blocks of size $8 \times 8$, which was obtained from generic $512 \times 512$ gray scale images. The $8 \times 8$ blocks are mapped to 1-D vectors, each of size $64$. The K-SVD removes the mean values of all blocks and reserves one dictionary element exclusively for the mean value. Although it is preferable that all elements except one have a zero mean, it is not a necessary for the design of dictionary. Thus, we preserved the mean
 values of all blocks. The initial dictionary for the K-SVD algorithm in Table \ref{table1} is the discrete cosine transform (DCT) of size $64\times 256$ with the sparsity level of K-SVD set at $64$. 

\begin{figure}[!htb]
\begin{center}
\includegraphics[width=.8\columnwidth]{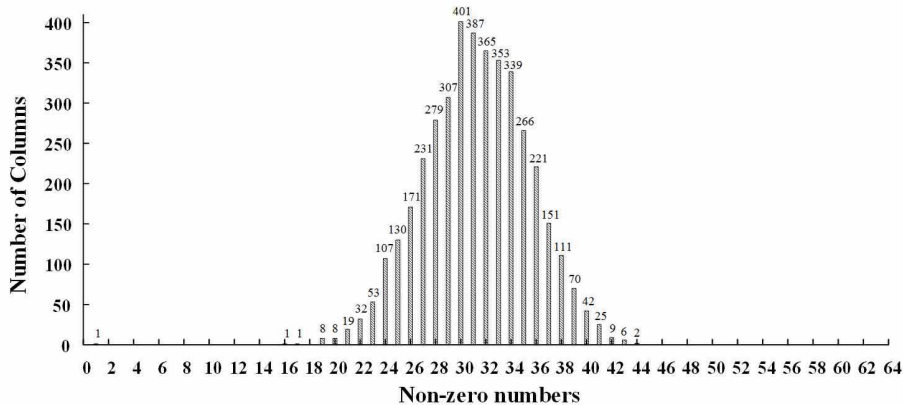}
\caption{Bell-shaped distribution of number of non-zero coefficients in columns of $X$ with a peak at $32$. Numerically, it is impossible to determine whether a number is zero; therefore, we set any number with an absolute value smaller than $10^{-6}$ as zero.}
\label{fig:non-zeros statistics}
\end{center}
\end{figure}

\begin{figure}[!ht]
\begin{center}
\includegraphics[width=.8\columnwidth]{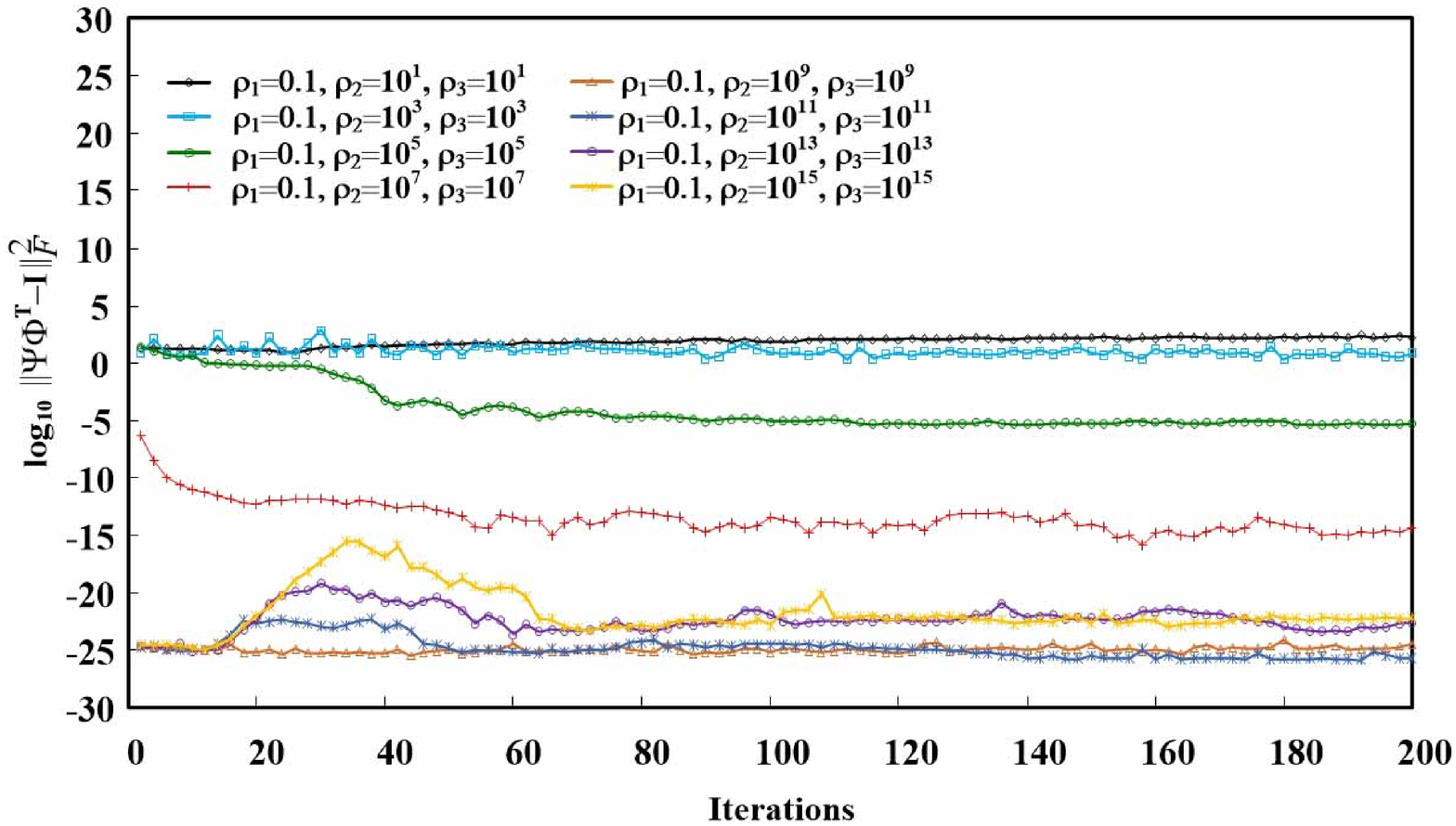}
\vspace{-0.05in}
\caption{Comparison of the convergence rate of $log_{10} (\left \|\psi\phi^\top -\mathcal I \right \|_F^2 )$ vs. number of iterations using various sets of $\rho_1$, $\rho_2$, and $\rho_3$ values. Curves corresponding to $\rho_2, \rho_3 \ge 10^{9}$ showed small numerical fluctuations of approximately $10^{-20}$ where number of iterations exceeded $100$. }
\label{fig:psiphiI}
\end{center}
\end{figure}

\begin{figure}[!ht]
\begin{center}
\includegraphics[width=.8\columnwidth]{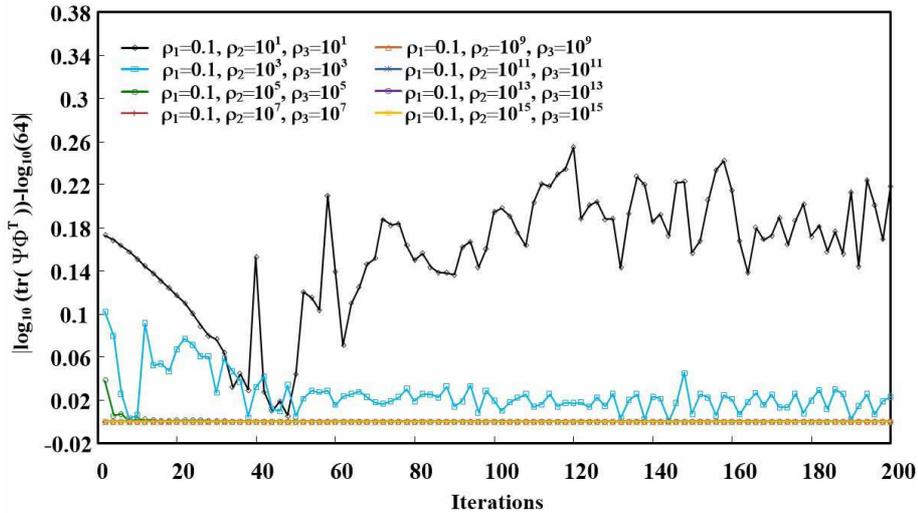}
\caption{Comparison of convergence rate of $\vert log_{10}(tr(\psi\phi^\top))-log_{10}(64)\vert$ vs. number of iterations using various sets of $\rho_1$, $\rho_2$, and $\rho_3$ values. Curves corresponding to $\rho_2, \rho_3 \ge 10^{7}$ showed small numerical fluctuations of approximately $0$ where number of iterations exceeded $100$. First term $\vert log_{10}(tr(\psi\phi^\top))-log_{10}(64)\vert$  measures the sum of the diagonal elements of $\psi\phi^\top$, the desired solution of which $64$. Thus, the value is subtracted from $\log_{10}(64)$. }
\label{fig:tracepsiphi}
\end{center}
\end{figure}

\begin{figure}[!ht]
\begin{center}
\includegraphics[width=.8\columnwidth]{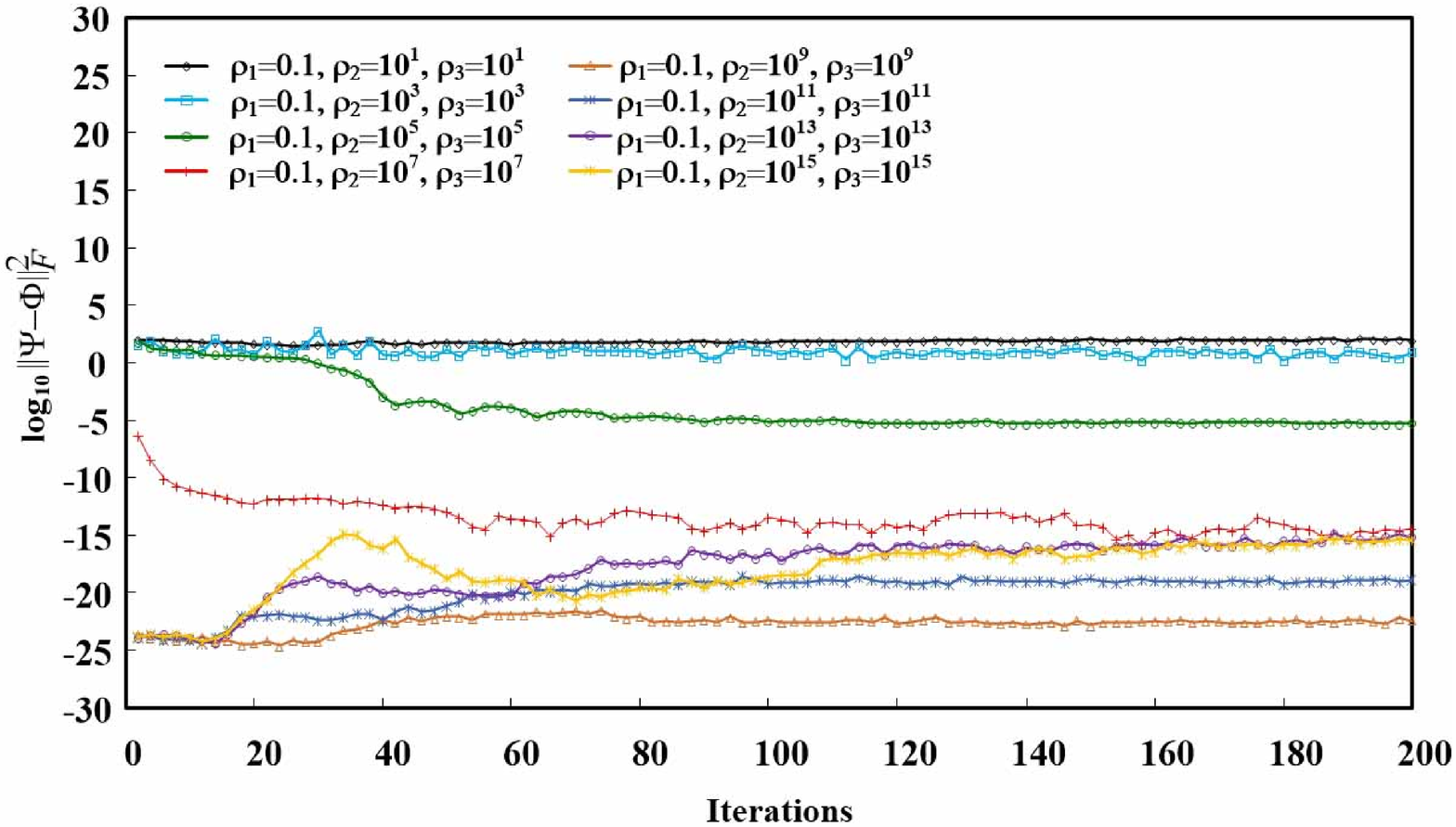}
\caption{Comparison of convergence rate of $log_{10} (\left \|\psi-\phi\right \|_F^2 )$ vs. number of iterations using various sets of $\rho_1$, $\rho_2$, and $\rho_3$ values. Curves corresponding to $\rho_2,\rho_3 \ge 10^{7}$ showed small numerical fluctuations of approximately $10^{-10}$ where number of iterations exceeded $100$. }
\label{fig:psiphi}
\end{center}
\end{figure}

\begin{figure}[!htb]
\begin{center}
\mbox{
\includegraphics[width=.22\columnwidth]{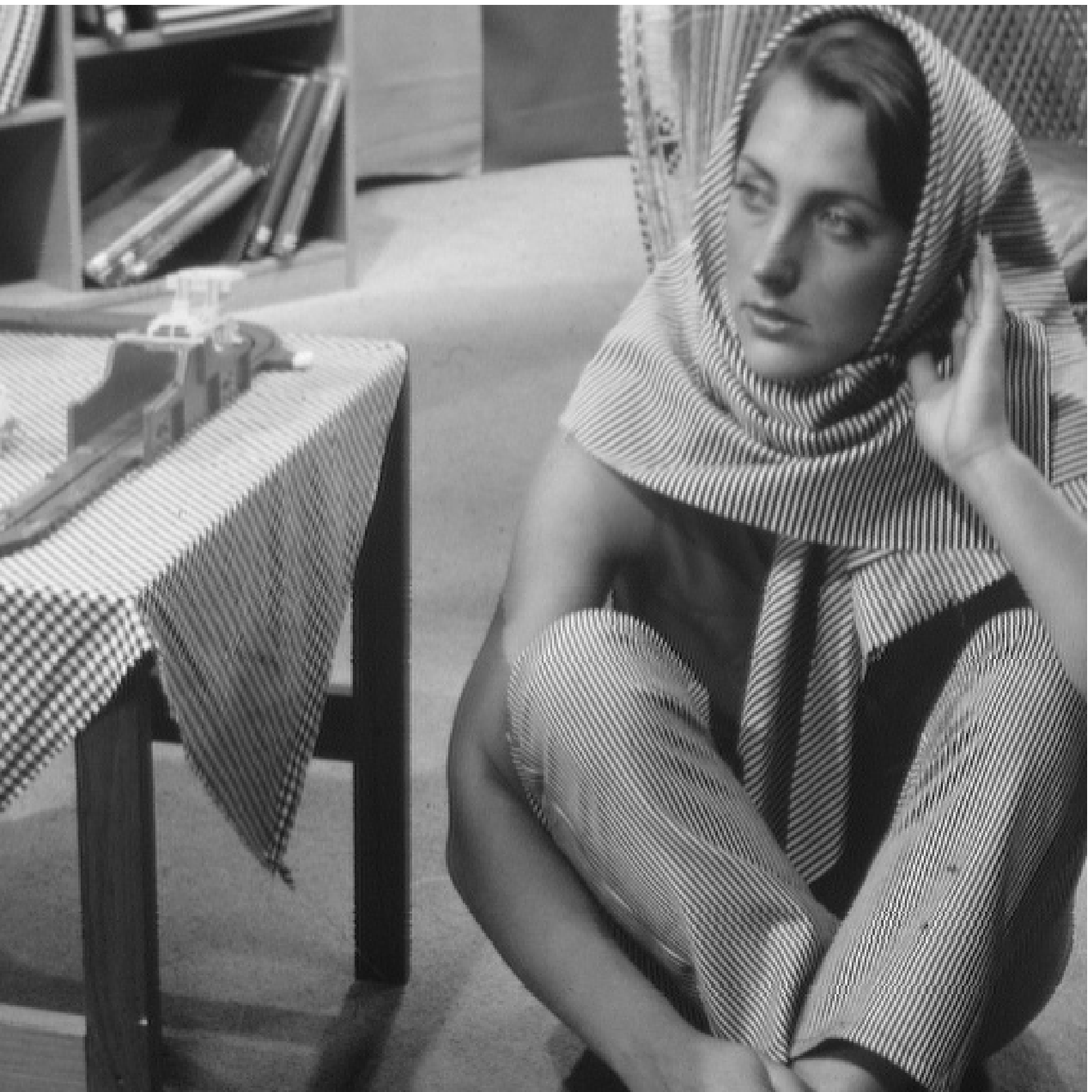}
}\\\medskip
\mbox{
\includegraphics[width=.22\columnwidth]{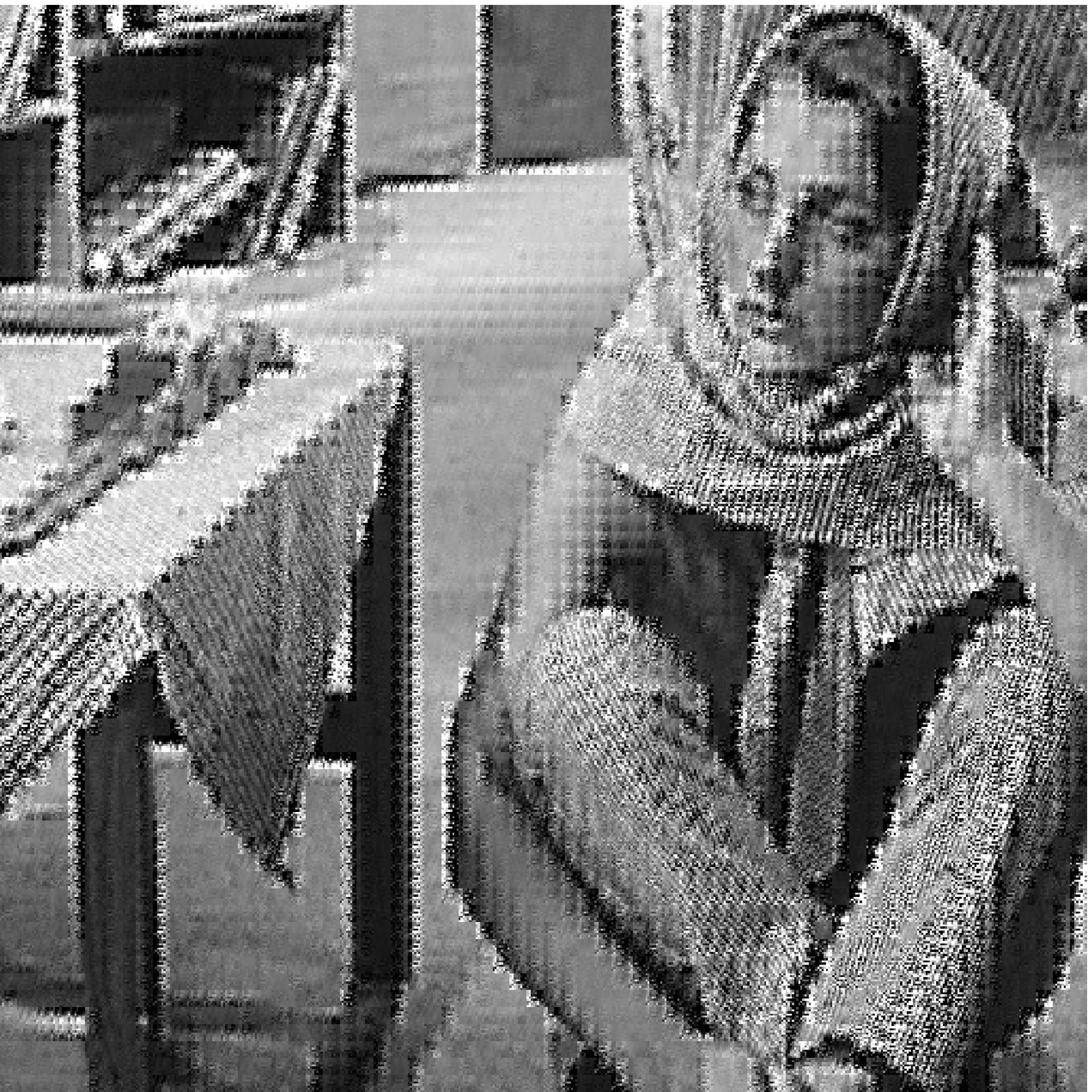}
\includegraphics[width=.22\columnwidth]{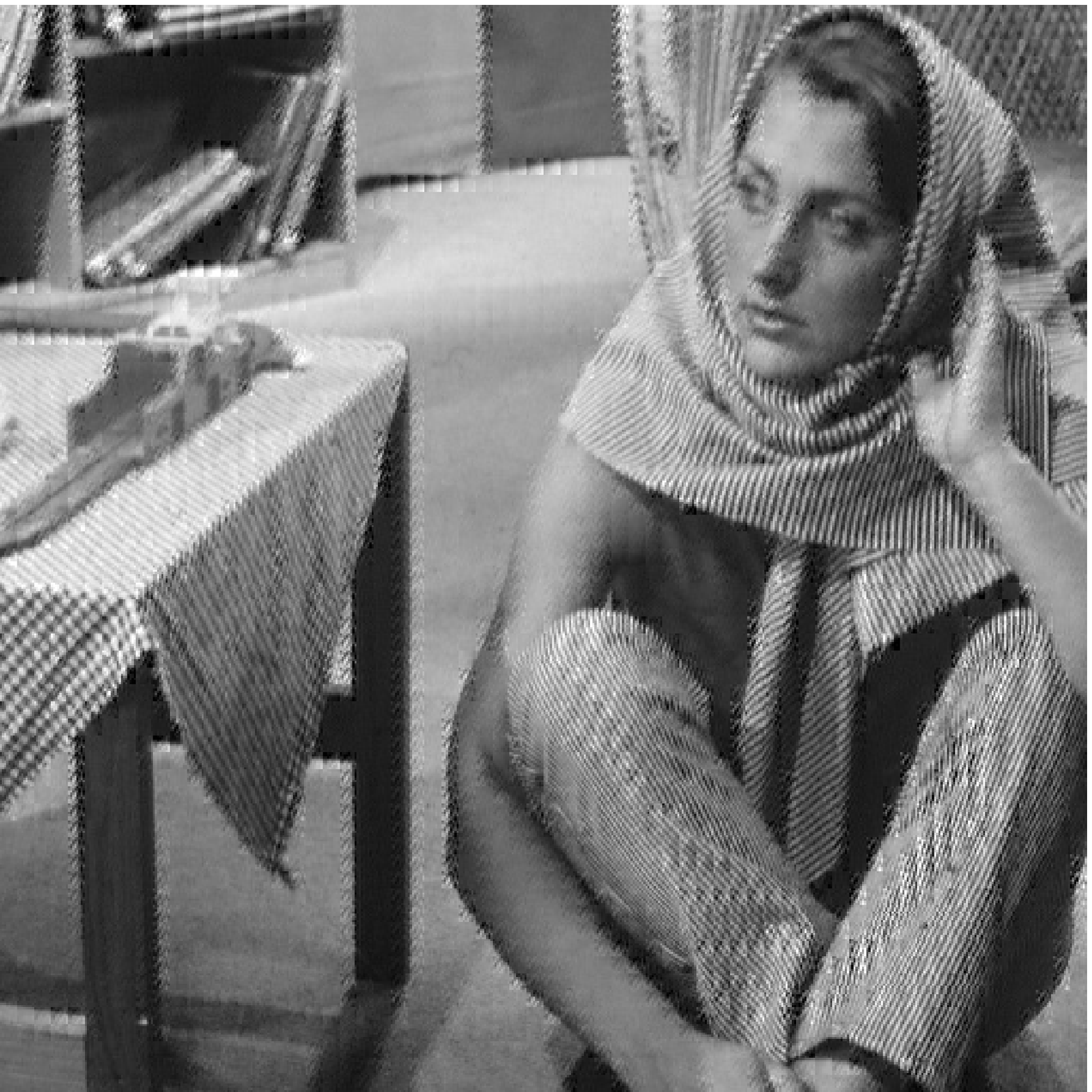}
\includegraphics[width=.22\columnwidth]{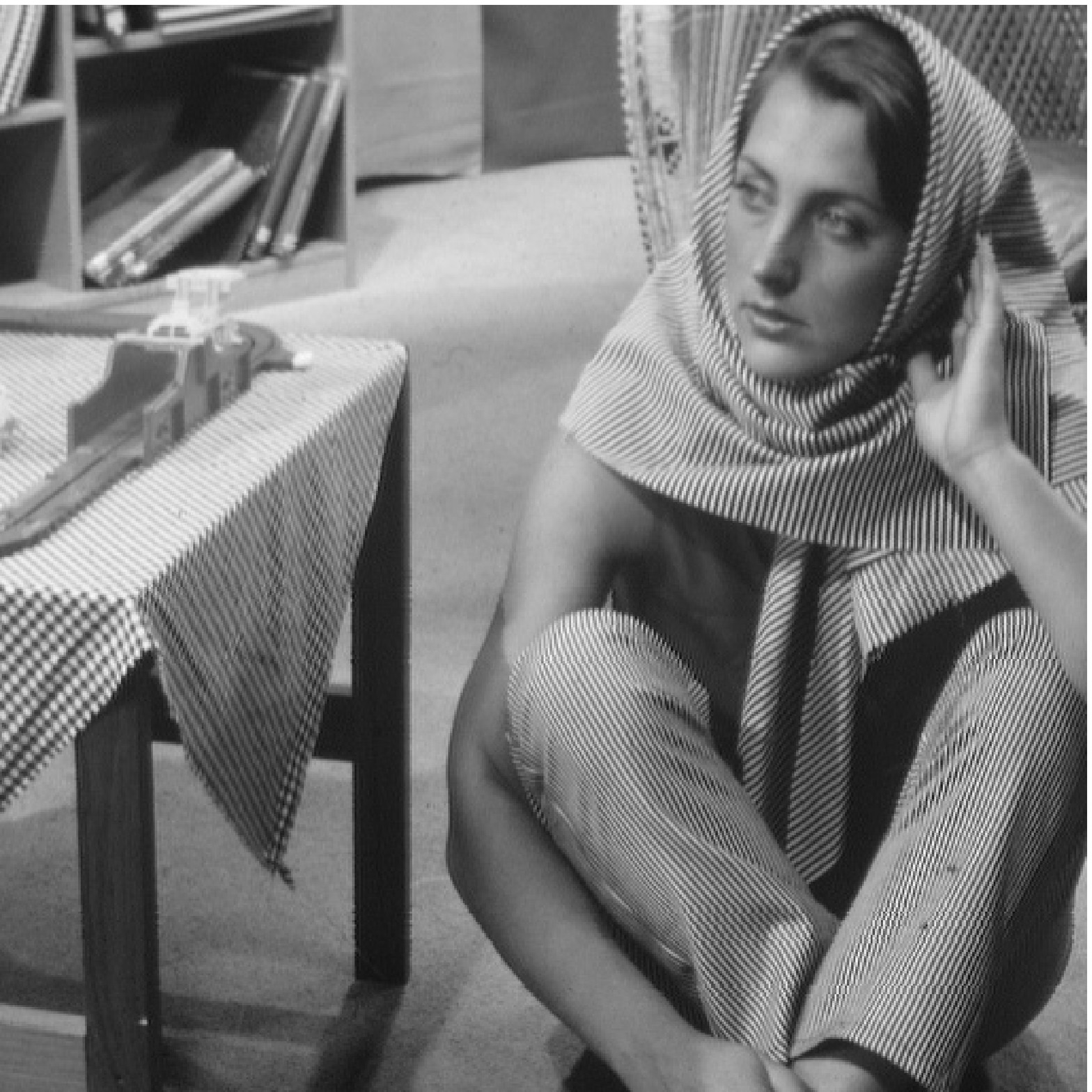}
\includegraphics[width=.22\columnwidth]{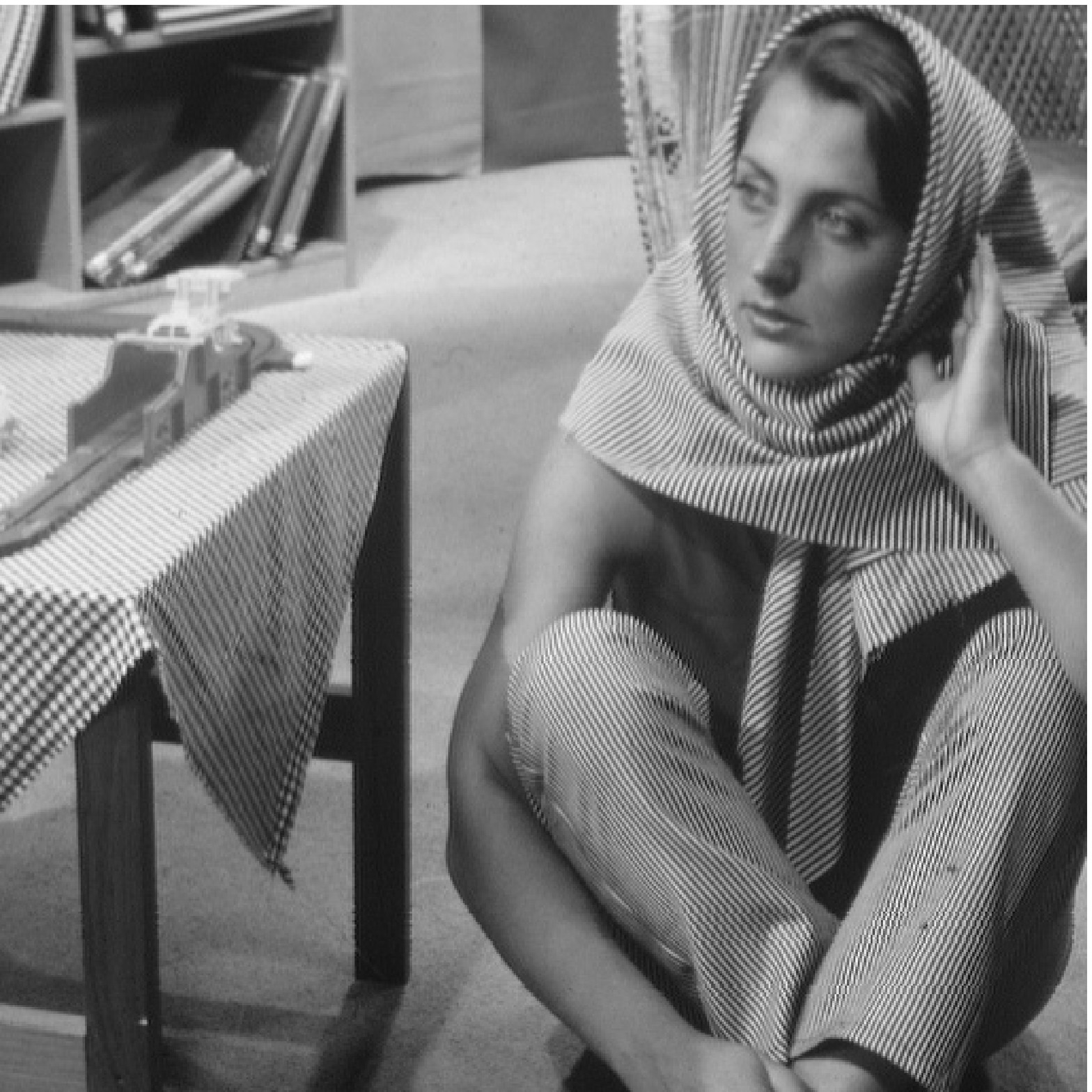}
}\\\medskip
\mbox{
\includegraphics[width=.22\columnwidth]{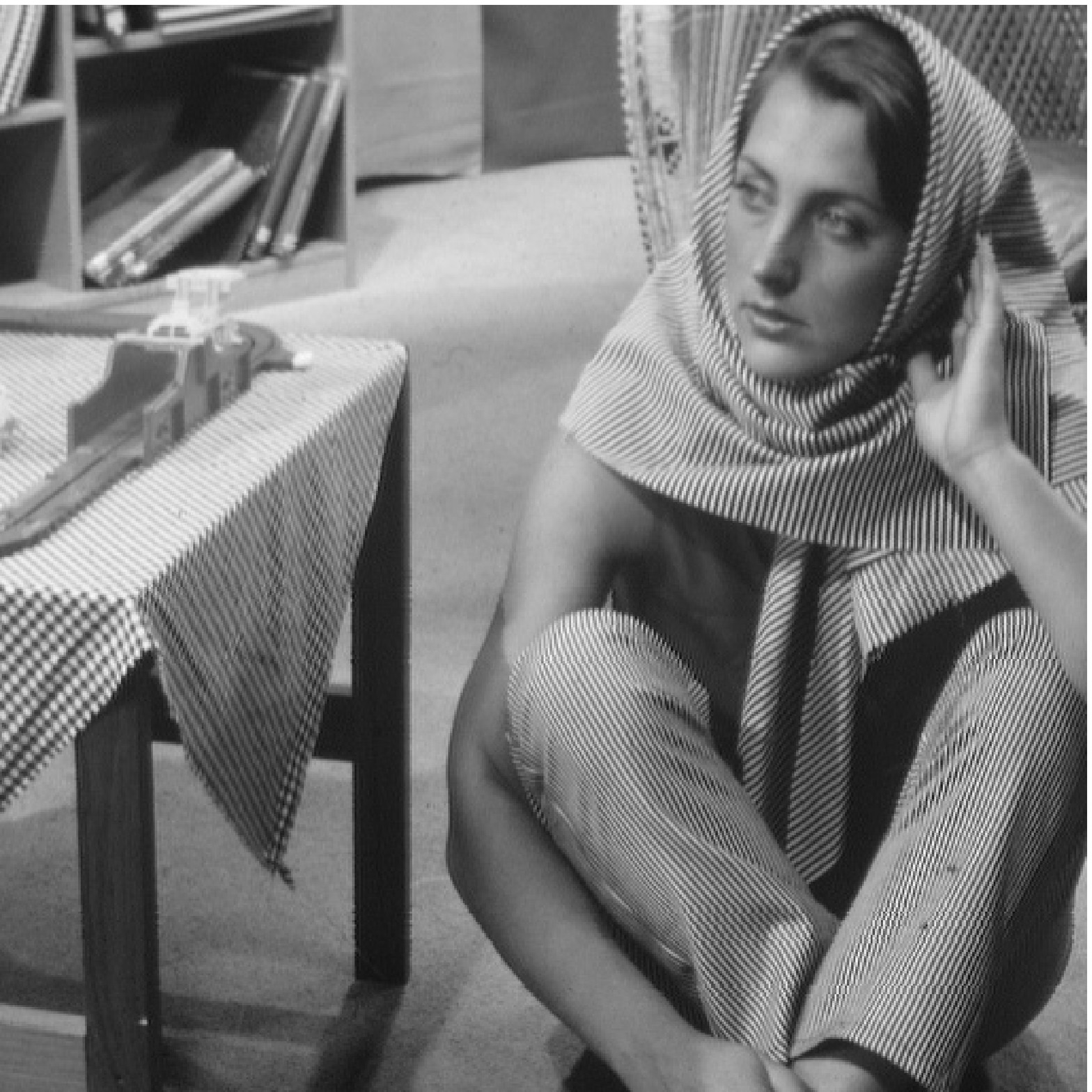}
\includegraphics[width=.22\columnwidth]{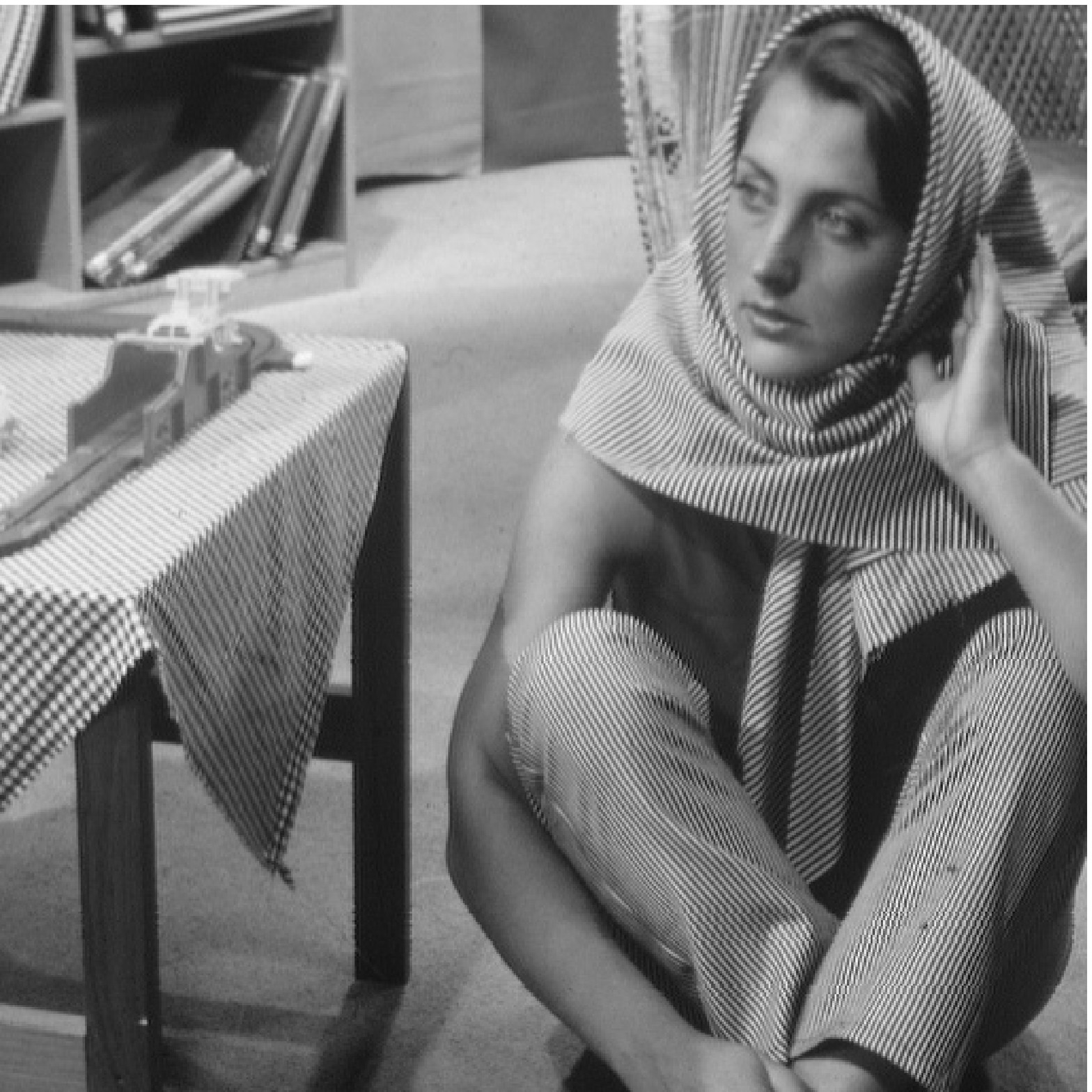}
\includegraphics[width=.22\columnwidth]{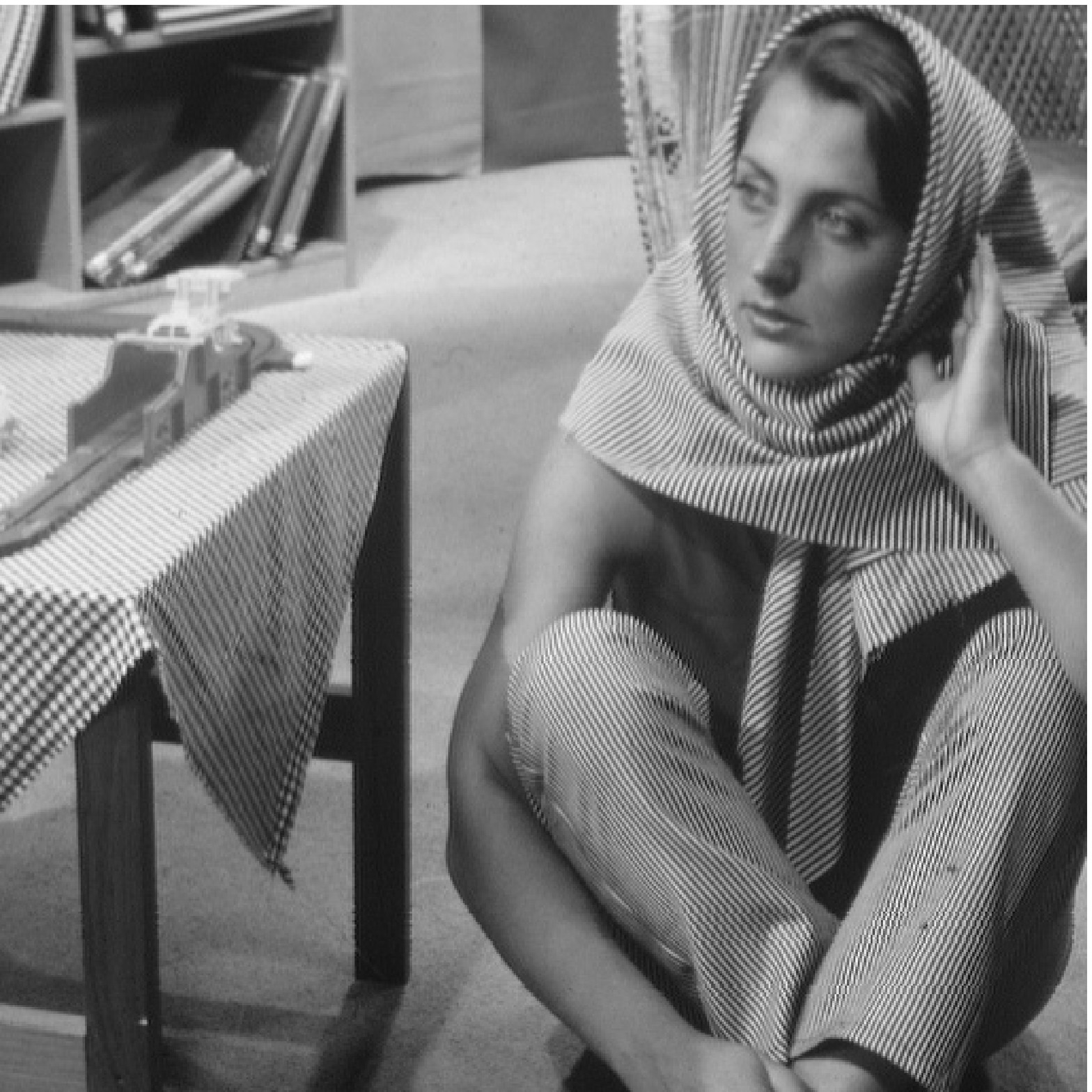}
\includegraphics[width=.22\columnwidth]{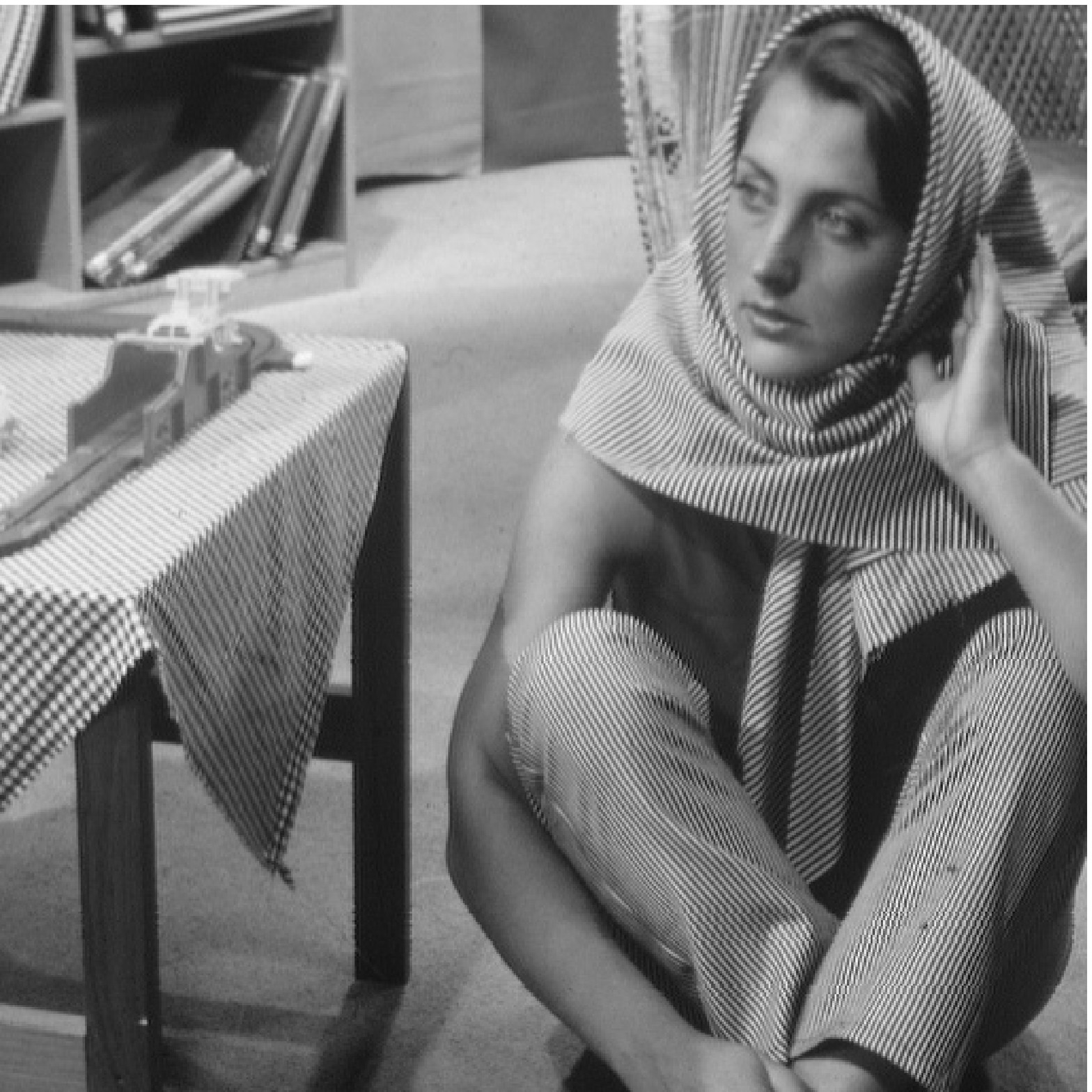}}
\caption{Reconstruction of image Barbara using flow diagram in Figure \ref{fig:reconstruction flow} with analysis and synthesis operations derived using values of $\rho_2$ and $\rho_3$. Top: Original image. 
Second row from left to right: Parameters are $\rho_2, \rho_3=10^1$, $\rho_2, \rho_3=10^3$, $\rho_2, \rho_3=10^5$ and $\rho_2, \rho_3=10^7$. Third row from left to right: Parameters are $\rho_2, \rho_3=10^9$, $\rho_2, \rho_3=10^{11}$, $\rho_2, \rho_3=10^{13}$ and $\rho_2, \rho_3=10^{15}$. Comparison of peak-signal-to-noise-ratio (PSNR) of reconstructed images: PSNRs of images in the second row are 16.09dB, 27.17dB, 88.27dB and 172.27dB, respectively; PSNRs of the images in the third row are 274.32dB, 288.01dB, 271.0dB and 266.51dB, respectively. The highest PSNR was from the image derived using $\rho_2 = \rho_3 = 10^{11}$ with $\rho_1$ fixed at $0.1$.}
\label{fig:DohmX_result_singl_image}
\end{center}
\end{figure}

\begin{figure}
\center
\includegraphics[width=.8\columnwidth]{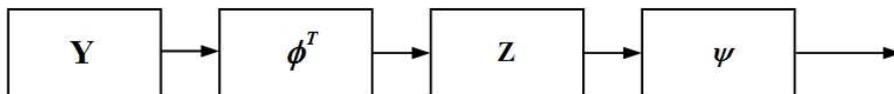}
\caption{Flow diagram of frame-based image reconstruction: Decomposition operation is followed by a reconstruction operation. An image is partitioned into non-overlapping blocks. Each block, a column in $Y$, is decomposed using analysis dictionary $\phi^\top$ to obtain the frame coefficients, a row of $Z$. The coefficients are then used to reconstruct the block using synthesis dictionary $\psi$.  
}
\label{fig:reconstruction flow}
\end{figure}

\begin{figure}[!ht]
\begin{center}
\includegraphics[width=.22\columnwidth]{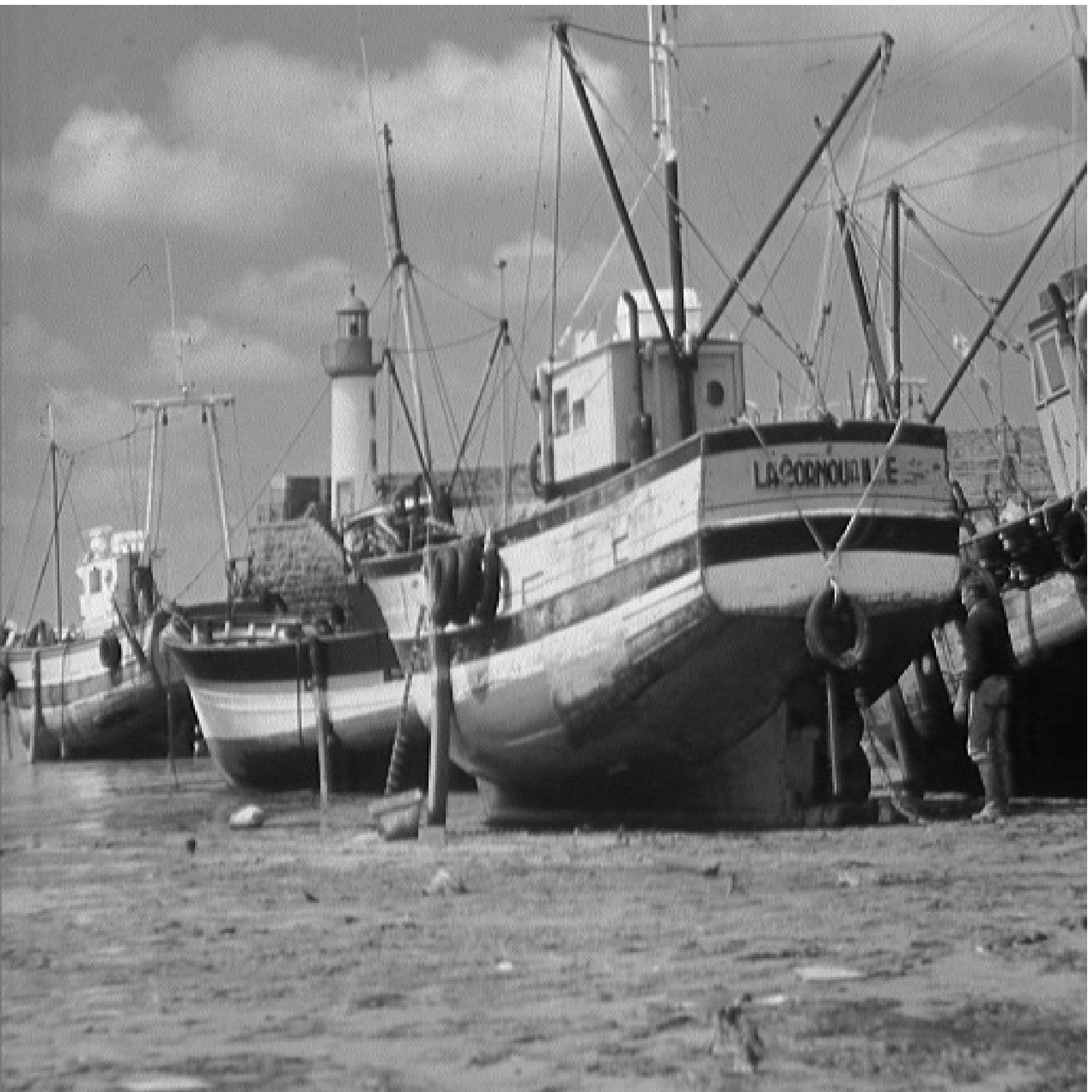}\ \ \ \ \ \ \ 
\includegraphics[width=.22\columnwidth]{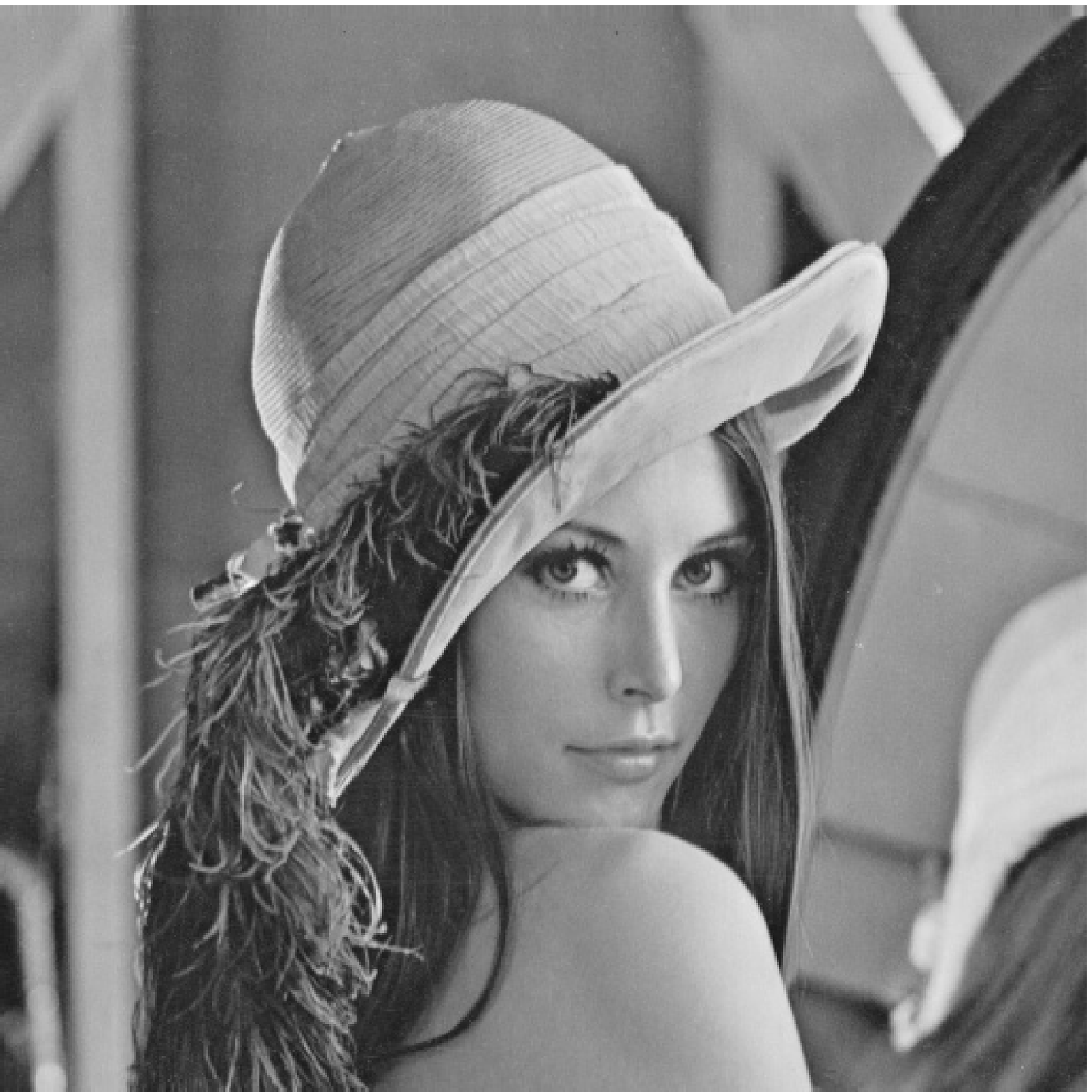}
\caption{Reconstruction of images Boat and Lena. The dictionaries were trained, using parameters $\rho_1 = 0.1$, $\rho_2=10^{11}$, and $\rho_3=10^{11}$, on image blocks taken from image Barbara. PSNRs of the reconstructed images Boat and Lena are $287.75$ dB and $289.20$ dB, respectively.}
\label{fig:DohmX_result_bar_train_lena_test_image}
\end{center}
\end{figure}

As previously discussed, the parameter $\rho_1$ is robust because its value does not change the minimizer of (\ref{objective0}) and is set to make $X$ update numerically stable by not dividing a zero in (\ref{equ:12}). Thus, $\rho_1$ is set at $0.1$.
We then applied the proposed algorithm to image blocks obtained from the image Barbara,
using various sets of parameters $\rho_2$ and $ \rho_3$ in conjunction with the matrices of Lagrangian multipliers $\lambda_2$ and $\lambda_3$, which were initially set at $0_{64 \times 64}$ and $0_{64\times 256}$, respectively. For each set of parameters, we conducted $l$ iterations and observed the convergence of functions corresponding to the equality constraints of (\ref{analysis0k-svd}) in terms of $log_{10} (\left \|\psi_l-\phi_l\right \|_F^2 )$, $log_{10} (\left \|\psi_l\phi_l^\top - \mathcal I \right \|_F^2 )$ and $\vert log_{10}(tr(\psi_l\phi_l^\top))-log_{10}(64)\vert$. The term $log_{10} (\left \|\psi_l-\phi_l\right \|_F^2 )$ measures the closeness of the constraint $\psi = \phi$ is reached and the last two terms measure that of the constraint $\psi_l\phi_l^\top = \mathcal I_{64 \times 64}$. Sparsity level $k$ in (\ref{analysis0k-svd}) for each column of $X$ is set at $64$. Figure \ref{fig:non-zeros statistics} illustrates the distribution of the number of non-zero coefficients in columns of $X$. 
The maximum number of iterations in Table \ref{table1} was set at $200$. As shown in Figures \ref{fig:psiphiI}, \ref{fig:tracepsiphi}, and \ref{fig:psiphi}, this is a sufficient  number of iterations to achieve numerical convergence for some parameter sets. As shown in the figures, large $\rho_2$ and $\rho_3$ values result in smaller objective values. Parameters $\rho_2$ and $\rho_3$ determine the severity of the penalty to the approximations of $\psi\phi^\top - \mathcal I$ and $\psi-\phi$, respectively, and the sizes of $\psi\phi^\top$ is $64 \times 64$ and that of $\psi$ is $64 \times 256$. 
The fact that $\rho_2$ and $\rho_3$ take higher values, means that the approximation become increasingly accurate.
Figure \ref{fig:DohmX_result_singl_image} displays the array of reconstructed images of Barbara, obtained through decomposition followed by  reconstruction operators using the learned dictionaries, as shown in Figure  \ref{fig:reconstruction flow}.  The quantity and quality performance depends on the values selected for $\rho_2$ and $\rho_3$. 
Experimental results indicate that respectively setting their values at $10^{11}$ and $10^{11}$ yields optimal PSNR performance in Figure \ref{fig:DohmX_result_singl_image}. This set of values is then fixed. Figure \ref{fig:DohmX_result_bar_train_lena_test_image} shows reconstructed images of Lena and Boat with the dictionaries trained using data from the image Barbara. 

\begin{figure}[!ht]
\begin{center}
\includegraphics[width=.5\columnwidth]{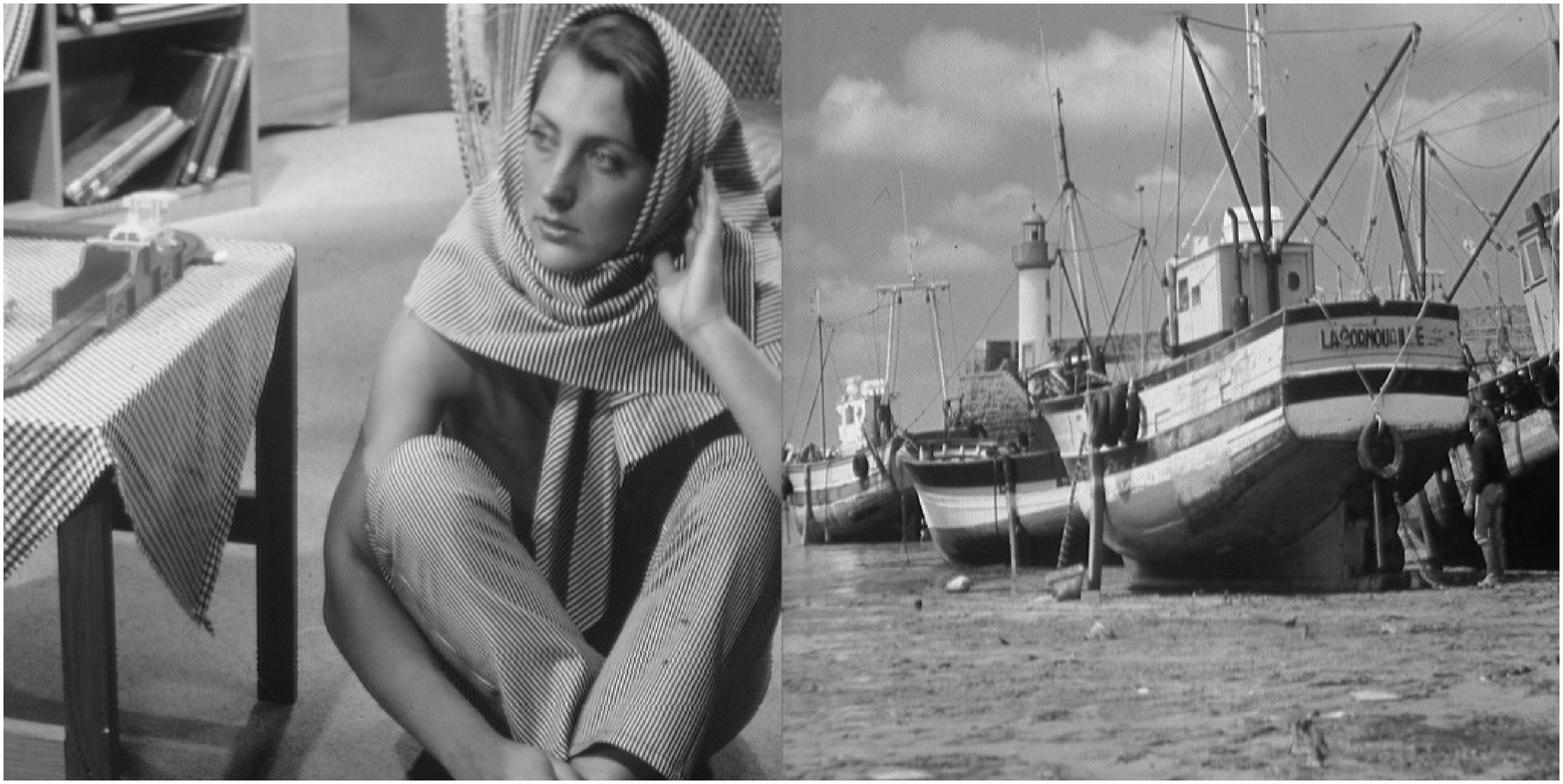}\ \ \ \ \ \ \ 
\includegraphics[width=.25\columnwidth]{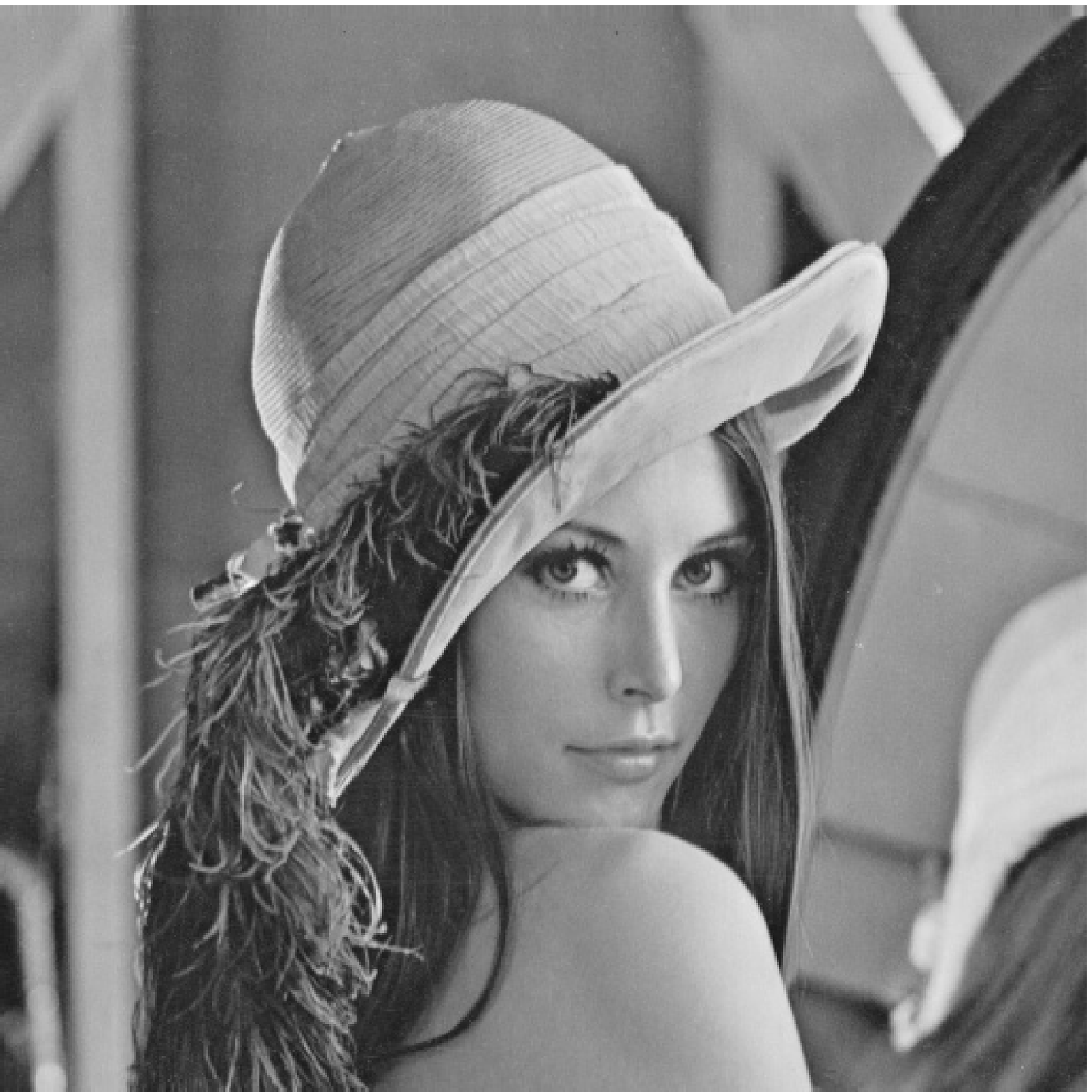}
\caption{Image Lena after reconstruction using dictionaries learned on image blocks of  Barbara and Boat, using parameters $\rho_1 = 0.1$, $\rho_2=10^{11}$, and $\rho_3=10^{11}$. The number of training blocks was $8,192$, which is double the number of training blocks used to derive the dictionaries for Figure \ref{fig:DohmX_result_bar_train_lena_test_image}.  PSNR of the reconstructed Lena is $288.19$ dB. By comparing the PSNR of Lena in Figure \ref{fig:DohmX_result_bar_train_lena_test_image}, we can see that increasing the number of training blocks does not improve the PSNR of reconstructed images.}
\label{fig:two_image_train}
\end{center}
\end{figure}

Figure \ref{fig:two_image_train} presents a reconstruction of the image Lena, which was derived using dictionaries learned using $8,192$ image blocks obtained from Barbara and Boat. This demonstrates that 
increasing the number of observations does not necessarily improve or quality of reconstructed images. 

\begin{figure}[!ht]
\begin{center}
\includegraphics[width=.25\columnwidth]{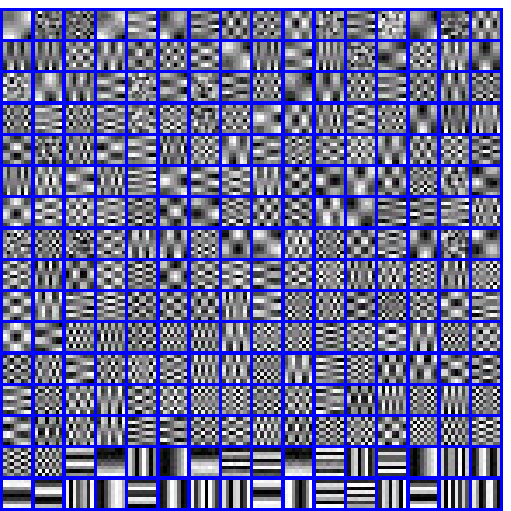} \ \
\includegraphics[width=.25\columnwidth]{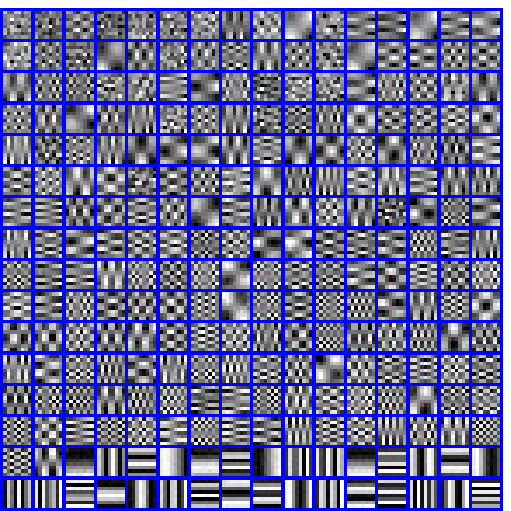}
\caption{Learned Parseval frame: (left)  trained on Barbara, (right) trained on Boat and Barbara. Columns of the frames are sorted in descending order of variance and stretched to their maximal range for the purpose of illustration. }
\label{fig:dict_images}
\end{center}
\end{figure}

\begin{figure}[!ht]
\begin{center}
\includegraphics[width=.9\columnwidth]{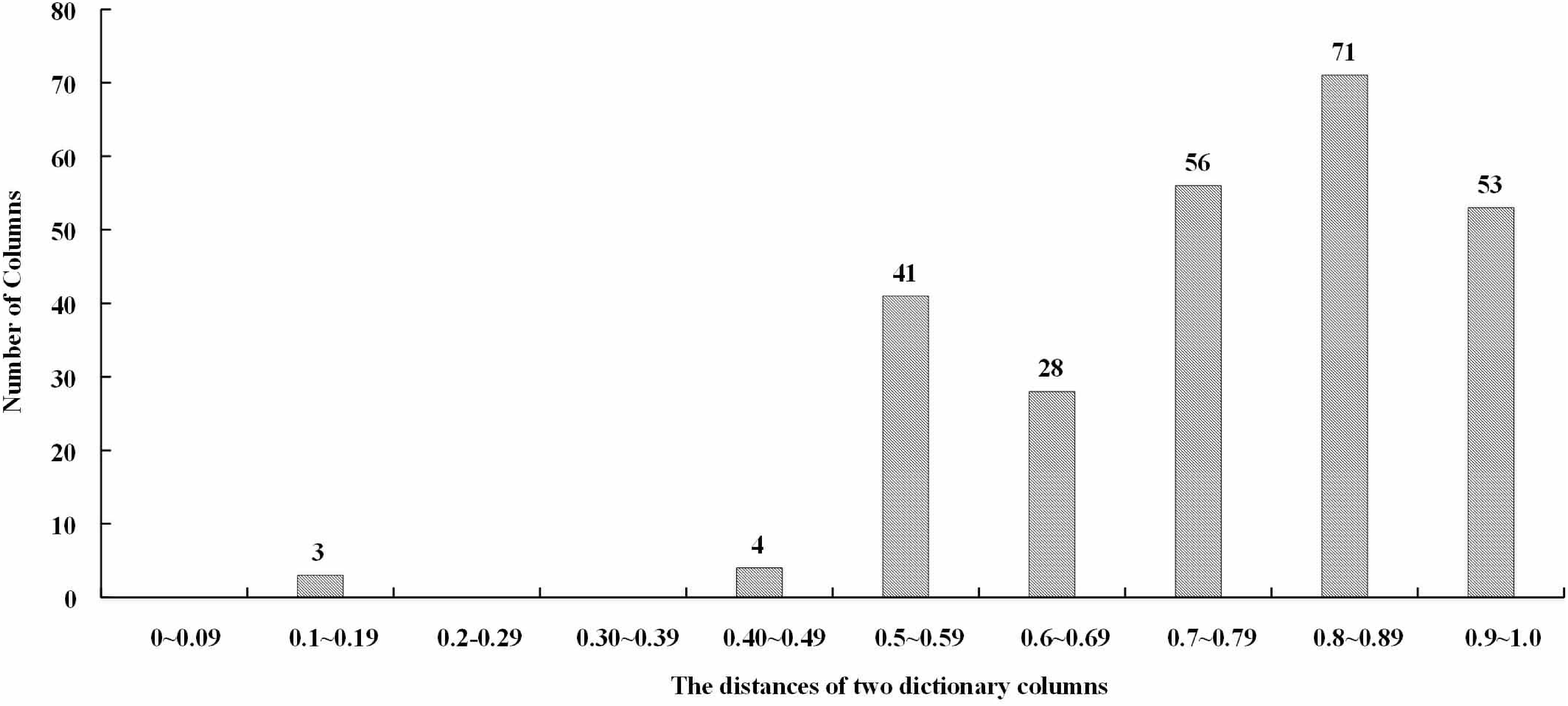}
\caption{Distribution of distances between elements  in the dictionary learned on Barbara and those in the dictionary learned on Barbara and Boat. (Most of the distances exceed $0.5$ and the range of distances is $[0, 1]$).}
\label{fig:dict_distances}
\end{center}
\end{figure}

The trained dictionaries are displayed in Figure \ref{fig:dict_images}. Low variance elements (bottom rows)  comprise separable horizontal and vertical cosine waves of different frequencies, whereas high variance elements (top rows) contain rotation-dependent variations and details. Similar patterns can be found in reports such as \cite{Lew00,Aharon06,Rub13}. Although they are perceptually similar, the two dictionaries are dissimilar in terms of quantity. Figure \ref{fig:dict_distances} shows the distribution of the distances of the $i$-th element $d_i$ between  dictionary $D$ and dictionary $E$, which is defined as follows:
\begin{equation}
1 - \max_j |d_i^\top e_j|,
\end{equation}
where $e_j$ is the $j$-th elements of dictionary $E$.

\vspace{5cm}




\subsection{Application to Image Processing: Preliminary results}

The optimization formulations using in the following experiments were designed to take advantage of the analysis and synthesis perspectives of representation coefficients (derived from frame theory) to problems involved in image recovery. Our objective in the performance comparison using the learned Parseval dictionary (the ratio of frame bounds is $1$) and the learned K-SVD dictionary (the ratio of frame bounds is $3.74$) was not to identity  the better dictionary, but rather to elucidate how dictionaries with different frame bounds would affect performance.

In denoising, tests were designed to leverage the sparse synthesis and redundant representations of an image. In image compression and filling-the-missing-pixels, we demonstrate that the effectiveness of a method relies on the frame bounds, which are used to obtain the redundancy
measure of the frame\cite{Pei97} in the learned frame dictionaries.

\hspace{3cm}

\subsubsection{Image Denoising} 

We address the following image denoising problem: 
\begin{eqnarray}
   Y = X+  N,
\label{equ:noise function}
\end{eqnarray}
where the ideal image $X$ is corrupted in the presence of additive white and homogeneous Gaussian noise, $N$, with mean zero and standard deviation of $\sigma$. 
We formulate this problem using the frame and the Bayesian approach as a constrained inverse problem. The objective of the optimization process is to use the learned dictionaries to recover the sparse synthesis coefficients $(W)$ of image with constrained redundant frame coefficients ($\phi^\top \psi W$).
\begin{eqnarray}
&\min  \limits_{W} \left\|W\right\|_1 \notag
\\
&{\rm subject \ to}\   \left\|Z- \phi^\top \psi W\right\|_2 \leq \varepsilon,
\label{equ:13}
\end{eqnarray}
where $\psi$ and $\phi$ are synthesis and analysis dictionaries, respectively; $Z$ is the frame coefficient of $Y$; and the value of $\varepsilon$ is related to $\sigma$. In implementation,  $\varepsilon$'s value is manually selected from a set of candidate value ranging from $2$ to $24$. The ones that yield the best performance are retained for use in deriving the denoised results from which the average performance is then measured.
This problem can be solved using the basis pursuit denoising (BPDN) method \cite{Chen01, Pat11}. 

Table \ref{tab:noise result} and Table \ref{tab:noise result SSIM} compare the average PSNR and structural similarities  (SSIM) between the images recovered using the proposed method with K-SVD and learned Parseval dictionaries
 in various noisy environments. 
As shown in the tables, using Parseval dictionaries obtains higher average PSNR and SSIM performance. In the experiments, the dictionary of K-SVD was used as the synthesis dictionary and its pseudo-inverse was adopted as the analysis dictionary. Note that the performance of the noisy results can be improved as long as structure in the frame coefficients can be explored\cite{Bal05, Elad06,Ho13}.

%

\begin{table}[ht]
\begin{center}
\caption{Comparison of average PSNR following noise reduction task. The means of the orignal images were subtracted from observations in experiments. The means were then added to the restored images from which PSNR values were obtained. The values are the averages obtained from five experiments using the same noise level ($\sigma$). Each block of a fixed $\sigma$ comprises four rows. The top row is the PSNR of a noisy image. The second row is the PSNR of an image processed using the K-SVD dictionaries.
The third row is the PSNR of an image processed using the learned Parseval dictionaries. The PSNR unit is $dB$.} \label{tab:noise result}
\begin{tabular}{|c|c||c||c||c||c||c||c|}
\hline 
\multicolumn{2}{|c||} {$\sigma$}&  Barbara & Man & Lena & Hill & Boat & Average\\\cline{1-8}
\hline\hline  & noisy image  & 34.16 & 34.16 & 34.16 & 34.16 & 34.16  &34.16\\
                  5 & K-SVD dictionaries &34.77 &34.73	&35.11	&34.73	&34.76	&34.82\\
                     & Parseval dictionaries &\bf{35.82} & \bf{35.78} & \bf{36.53} & \bf{35.69} & \bf{35.73} &\bf{35.91} \\\cline{1-8}                
\hline\hline  &  noisy image &28.14 & 28.14 & 28.14 & 28.14 & 28.15 &28.14\\
                10 & K-SVD dictionaries &29.82 &29.98	&31.01	&30.22	&30.14	&30.23\\
                    & Parseval dictionaries &\bf{31.22} & \bf{31.34} & \bf{32.53} & \bf{31.44} & \bf{31.41} & \bf{31.59}\\\cline{1-8}                                
\hline\hline  & noisy image &24.64 & 24.62 & 24.62 & 24.64 & 24.65 & 24.63\\
                15 & K-SVD dictionaries &27.37	&27.82	&28.98	&28.20	&27.95	&28.07\\ 
                     & Parseval dictionaries &\bf{28.78} & \bf{29.10} & \bf{30.44} & \bf{29.34} & \bf{29.21} & \bf{29.37}\\\cline{1-8}                                              
\hline\hline  & noisy image &22.18 & 22.15 & 22.14 & 22.17 & 22.19 &22.17\\
           20 & K-SVD dictionaries &25.82 &26.57	&27.79	&27.04	&26.61	&26.77\\
              & Parseval dictionaries &\bf{27.17} & \bf{27.69} & \bf{29.06} & \bf{28.02} & \bf{27.78} &\bf{27.94}\\\cline{1-8}                       
\hline\hline  & noisy image &20.31 & 20.25 & 20.24 & 20.29 & 20.29 &20.28\\
		   25 & K-SVD dictionaries &24.73 &25.66	&26.83	&26.16	&25.62	&25.80\\
              & Parseval dictionaries &\bf{26.00} & \bf{26.70} & \bf{28.08} & \bf{27.11} & \bf{26.75} &\bf{26.93}\\\cline{1-8}                                  
\hline\hline  & noisy image &18.80 & 18.73 & 18.71 & 18.76 & 18.75 &18.75\\
           30 & K-SVD dictionaries &23.92 &25.07	&26.25	&25.59	&24.92	&25.15\\  
              & Parseval dictionaries &\bf{25.11} &\bf{ 25.93} &\bf{ 27.24} & \bf{26.39} & \bf{25.94} &\bf{26.12}\\\cline{1-8}               
\end{tabular}
\end{center}
\end{table}

\begin{table}[!hb]
\begin{center}
\caption{Comparison of average SSIM associated with noise reduction task. The means of the orignal images were subtracted from observations in experiments. The means were then added to the restored images from which SSIM values were obtained. The values are the average obtained from five experiments using the same noise level ($\sigma$). Each block of a fixed $\sigma$ consists of four rows. The top row is the SSIM of a noisy image. The second row is the SSIM of an image processed using the K-SVD dictionaries. 
The third row is the SSIM of an image processed using learned Parseval dictionaries.} \label{tab:noise result SSIM}
\begin{tabular}{|c|c||c||c||c||c||c||c|}
\hline 
\multicolumn{2}{|c||} {$\sigma$}&  Barbara & Man & Lena & Hill & Boat &Average\\\cline{1-8}
\hline\hline    & noisy image &0.89	&0.88 &0.85	&0.89 &0.89	&0.88\\
              5 & K-SVD dictionaries &0.91 &0.90 &0.87 &0.90 &0.90 &0.90\\               
                & Parseval dictionaries &\bf{0.94}	&\bf{0.92} &\bf{0.91} &\bf{0.92} &\bf{0.92} &\bf{0.92}\\\cline{1-8}               
\hline\hline    & noisy image &0.72	&0.68 &0.61 &0.69 &0.69 &0.68\\
			 10 & K-SVD dictionaries &0.81 &0.77 &0.76 &0.78 &0.78 &0.78\\
                & Parseval dictionaries &\bf{0.86}	&\bf{0.83} &\bf{0.83} &\bf{0.82} &\bf{0.83} &\bf{0.83}\\\cline{1-8}                 
\hline\hline    & noisy image &0.58	&0.52 &0.45 &0.53 &0.54 &0.52\\
             15 & K-SVD dictionaries &0.74 &0.69	&0.69 &0.69	&0.70 &0.70\\
                & Parseval dictionaries &\bf{0.80}	&\bf{0.75} &\bf{0.77}	&\bf{0.74} &\bf{0.76}	&\bf{0.76}\\\cline{1-8}                               
\hline\hline    & noisy image &0.48	&0.41 &0.34	&0.41 &0.43	&0.42\\
             20 & K-SVD dictionaries &0.68 &0.63	&0.66 &0.63	&0.65 &0.65\\
                & Parseval dictionaries&\bf{0.74}	&\bf{0.69} &\bf{0.72}	&\bf{0.68} &\bf{0.70} &\bf{0.71}\\\cline{1-8}                      
\hline\hline    & noisy image &0.40 &0.33 &0.27 &0.33 &0.35 &0.34\\
             25 & K-SVD dictionaries &0.63 &0.59	&0.62 &0.58	&0.60 &0.60\\
                & Parseval dictionaries &\bf{0.70}	&\bf{0.65} &\bf{0.69}	&\bf{0.63} &\bf{0.66}	&\bf{0.67}\\\cline{1-8}                              
\hline\hline    & noisy image &0.35 &0.27 &0.22 &0.27 &0.29 &0.28\\
             30 & K-SVD dictionaries &0.59 &0.57 &0.61 &0.55 &0.57 &0.58\\
                & Parseval dictionaries &\bf{0.66}	&\bf{0.61} &\bf{0.66}	&\bf{0.59} &\bf{0.62}	&\bf{0.63}\\\cline{1-8}                 	 		
\end{tabular}
\end{center}
\end{table}

\subsubsection{Image Compression}

\begin{figure}
\begin{center}
\includegraphics[width=.8\columnwidth]{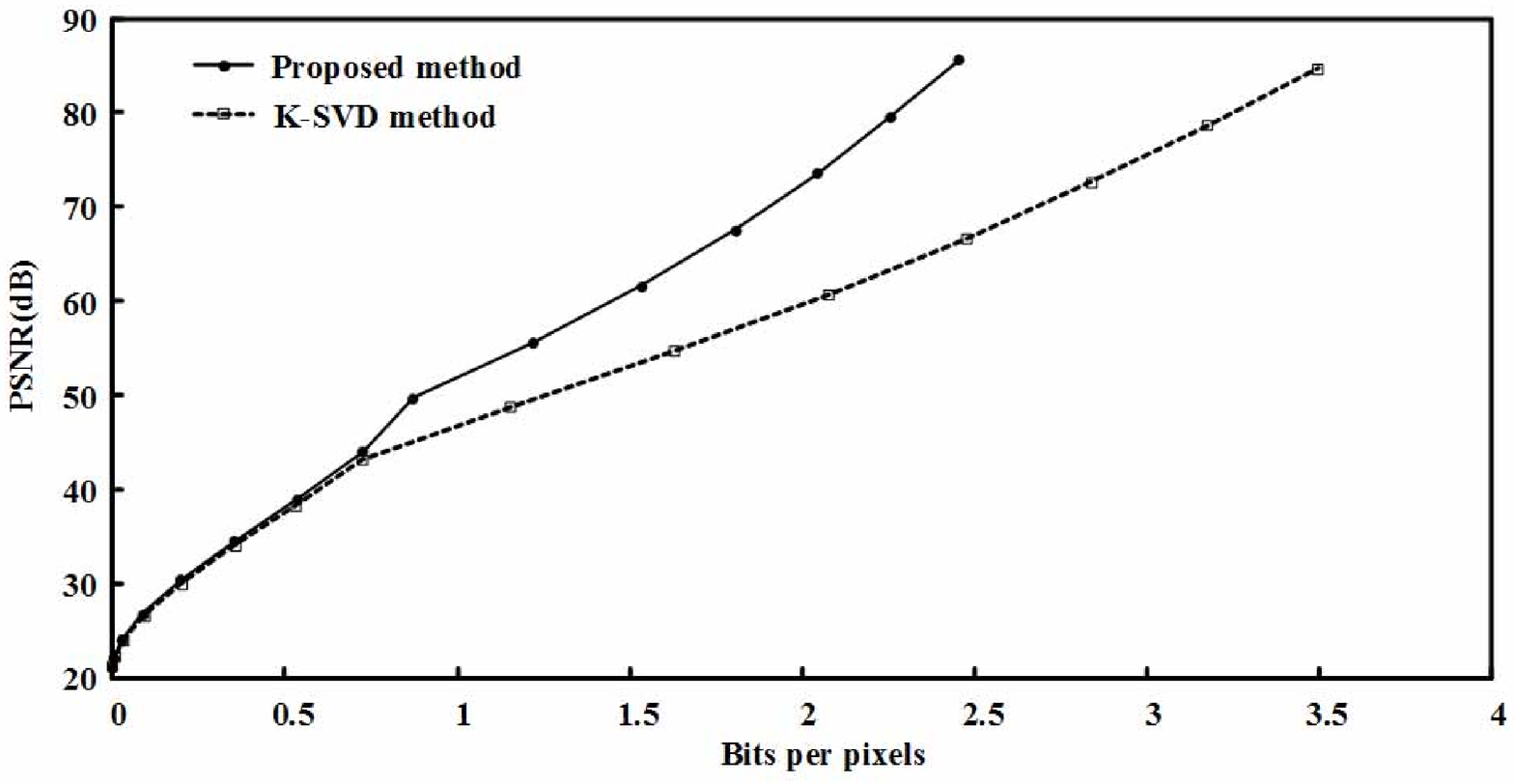}
\includegraphics[width=.8\columnwidth]{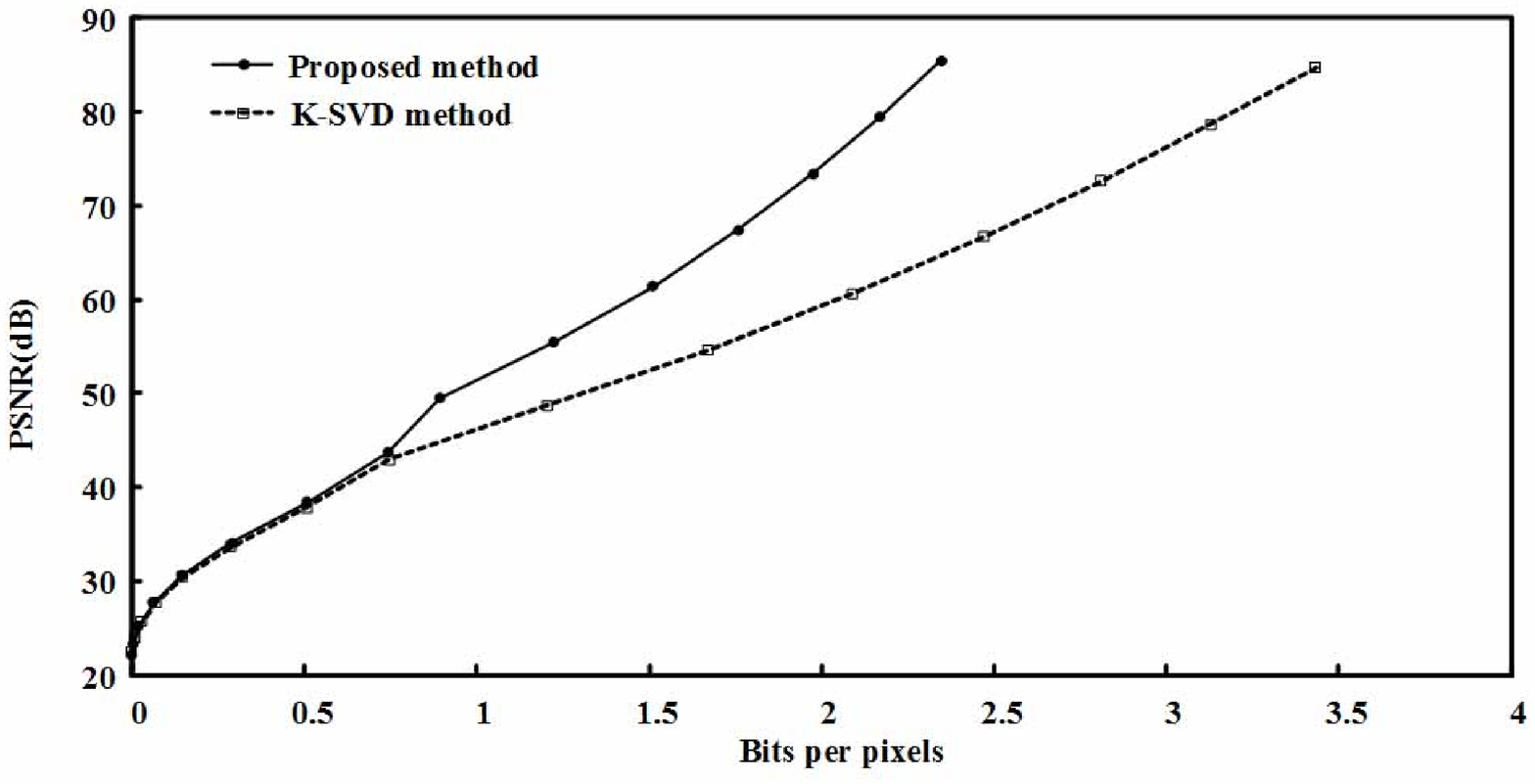}
\caption{Rate-distortion graphs of images: Barbara (Top) and Boat (Bottom). Proposed method uses the learned Parseval analysis dictionary to obtain frame coefficients, whereas the K-SVD method is based on the pseudo-inverse of the K-SVD dictionary.}
\label{fig:123}
\end{center}
\end{figure}

We conducted a comparison of image compression between the synthesis dictionary learned using the K-SVD method and its pseudo-inverse as the analysis dictionary and the proposed Parseval dictionaries. The entropy of the bit distribution at each bit-plane of frame coefficients is  encoded under the assumption that the bit distribution is an independently identically distributed (i.i.d.) random variable. The resulting rate-distortion graph is presented in Figure \ref{fig:123}. An image was partitioned into $8 \times 8$ disjoint blocks. All blocks were encoded independently from the other blocks. The mean of the image was assumed to be known and subtracted from the image in the experiments. The mean was then added to the compressed image to measure the PSNR. As shown in Figure \ref{fig:123}, using the Parseval dictionary obtains a better performance than using the K-SVD dictionary. For a fixed PSNR, using the Parseval dictionary achieved a savings in bits per pixels (bpp) up to $0.7$ dB when its value was above $0.8$. 

The frame bounds can be used as an indication of the correlation between frame coefficients, wherein a lower value $\frac{B}{A}$ indicates lower correlation between frame coefficients\cite{Gro93}.
The above two experiments demonstrate the advantage of using a Parseval tight frame to remove the correlation between frame coefficients. The frame bounds $A$ and $B$ of the learned Parseval frame are $1$ whereas those of K-SVD are respectively $1$ and $3.74$. The smaller $\frac{B}{A}$ value of a Parseval frame is the reason why the method based on Parserval dictionaries obtains superior performance in image compression. Similarly, a high degree of correlation between frame coefficients facilitates the recovery of missing pixels, due to the fact that missing coefficients can be compensated for using other coefficients through high frame redundancy.

\subsubsection{Filling Missing Pixels} 

Information pertaining to an image is given by a sequence of irregularly sampled pixels. This means that the problem lies in determining the means by which to recover the image from the input. 

We partitioned an image into $8 \times 8$ blocks. We then randomly removed a fraction of the pixels of an image block (between $0.1$ and $0.8$) by setting their values at zero. 
Let $X$ denote the ideal image and let $Q$ be a matrix of either $0$ or $1$, respectively indicating the positions of the corrupted and non-corrupted pixels in $X$. Thus, their product, $QX$, consists exclusively of non-corrupted pixels in $X$. For example, if $X_k$=[$2$ $3$ $5$ $0$ $0$ $1$]$^\top $ is the $k$-th column of $X$, then the $k$-th column of $QX$ is [$2$ $3$ $5$ $1$]$^\top $.

We formulate the problem using the frame as a constrained inverse problem, with the aim of recovering the sparse synthesis coefficients ($W$) of image $X$: 
\begin{eqnarray}
&\min  \limits_{W} \left\|W\right\|_1 \notag
\\
&{\rm subject \ to}\   \left\|QX-Q \psi W\right\|_2 \leq \varepsilon,
\label{equ:13}
\end{eqnarray}
where $\varepsilon$ is a parameter set at $0.01$ for all cases; and $\psi$ is the synthesis dictionary. This problem can be solved using the BPDN algorithm. The reconstructed image is obtained by applying synthesis frame $\psi$ to solution $W$. 

\begin{figure}[!h]
\begin{center}
\centering
\mbox{
\includegraphics[width=0.25\textwidth]{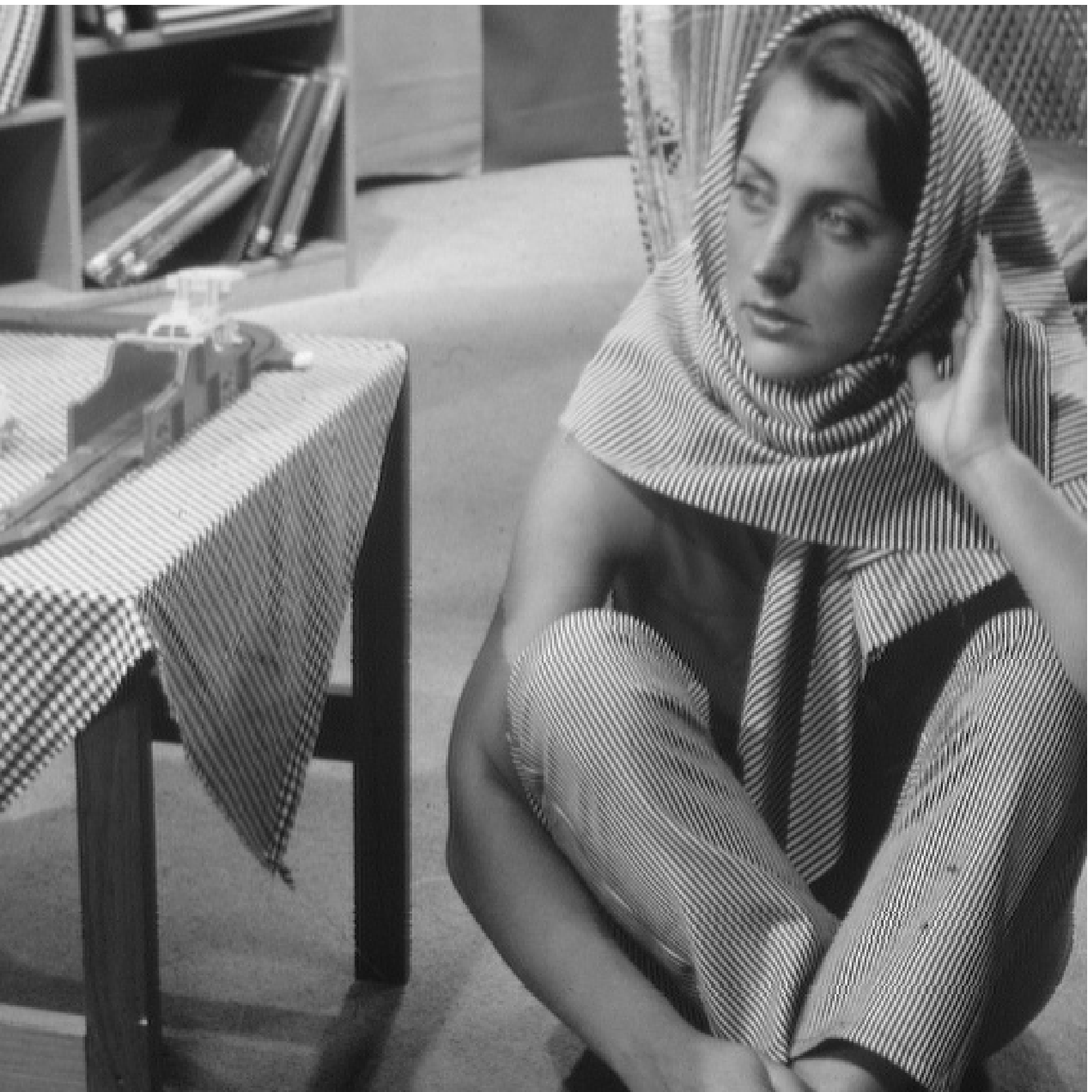}}\\\medskip
\centering
\mbox{
\includegraphics[width=0.25\textwidth]{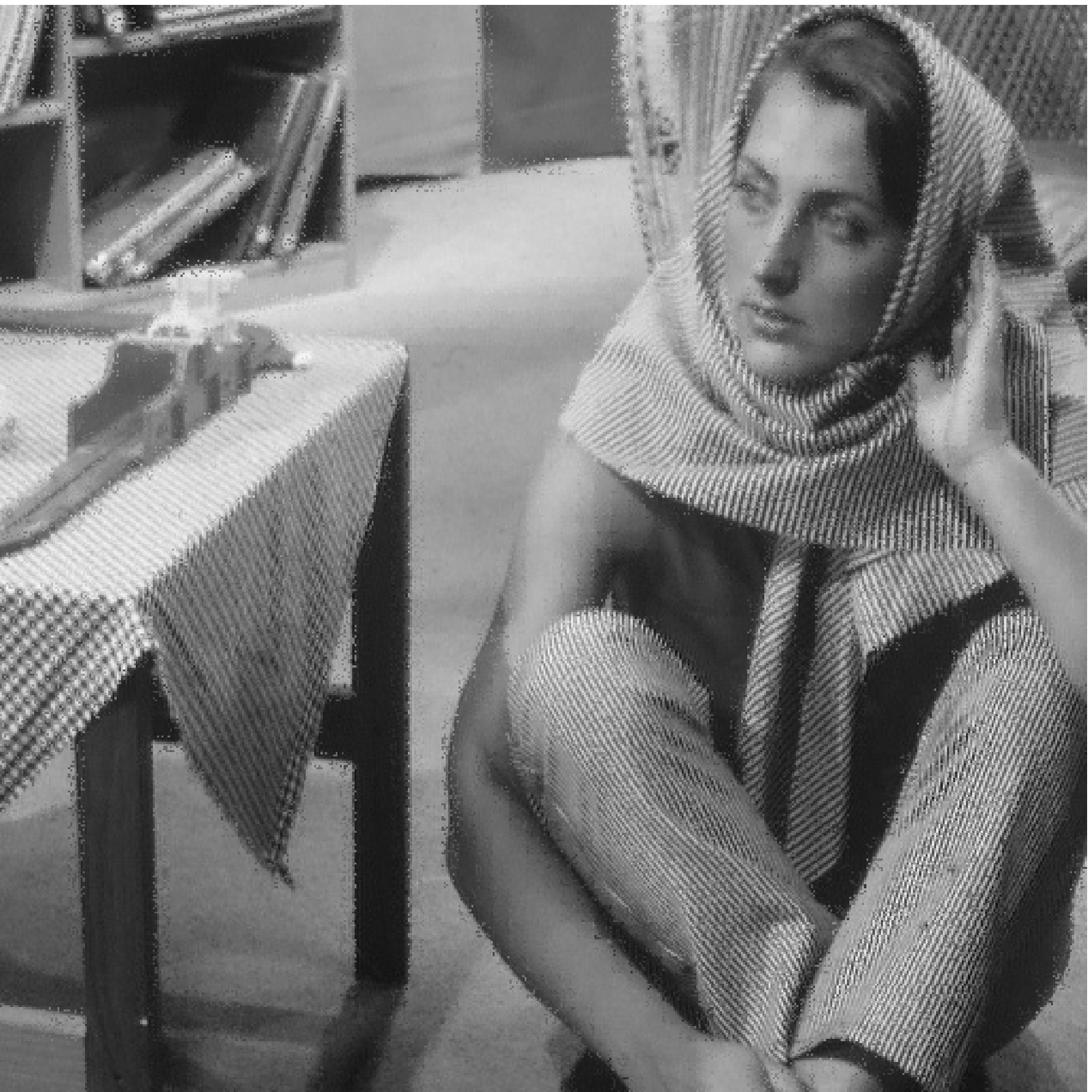}
\includegraphics[width=0.25\textwidth]{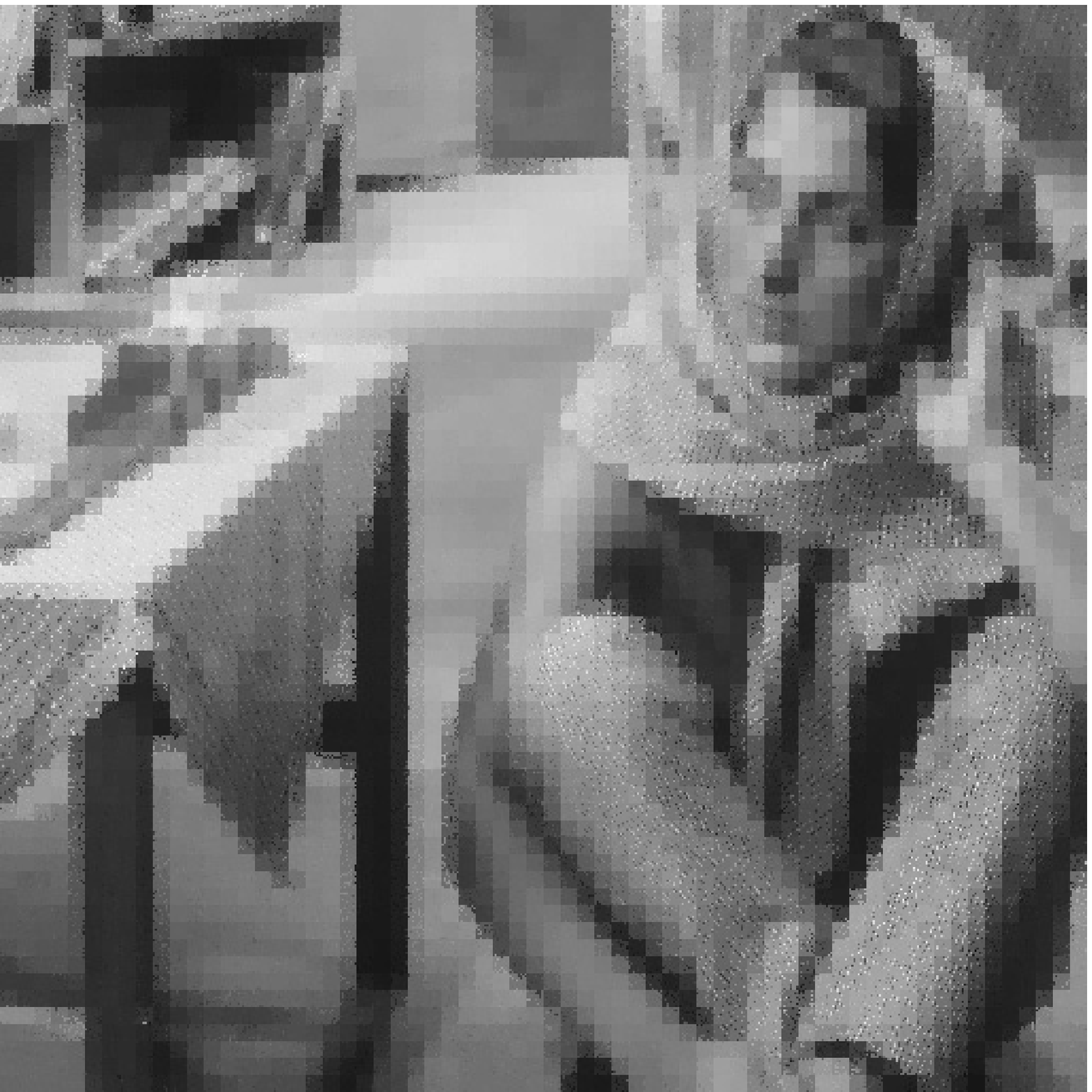}
}\\\medskip  
\mbox{
\includegraphics[width=0.25\textwidth]{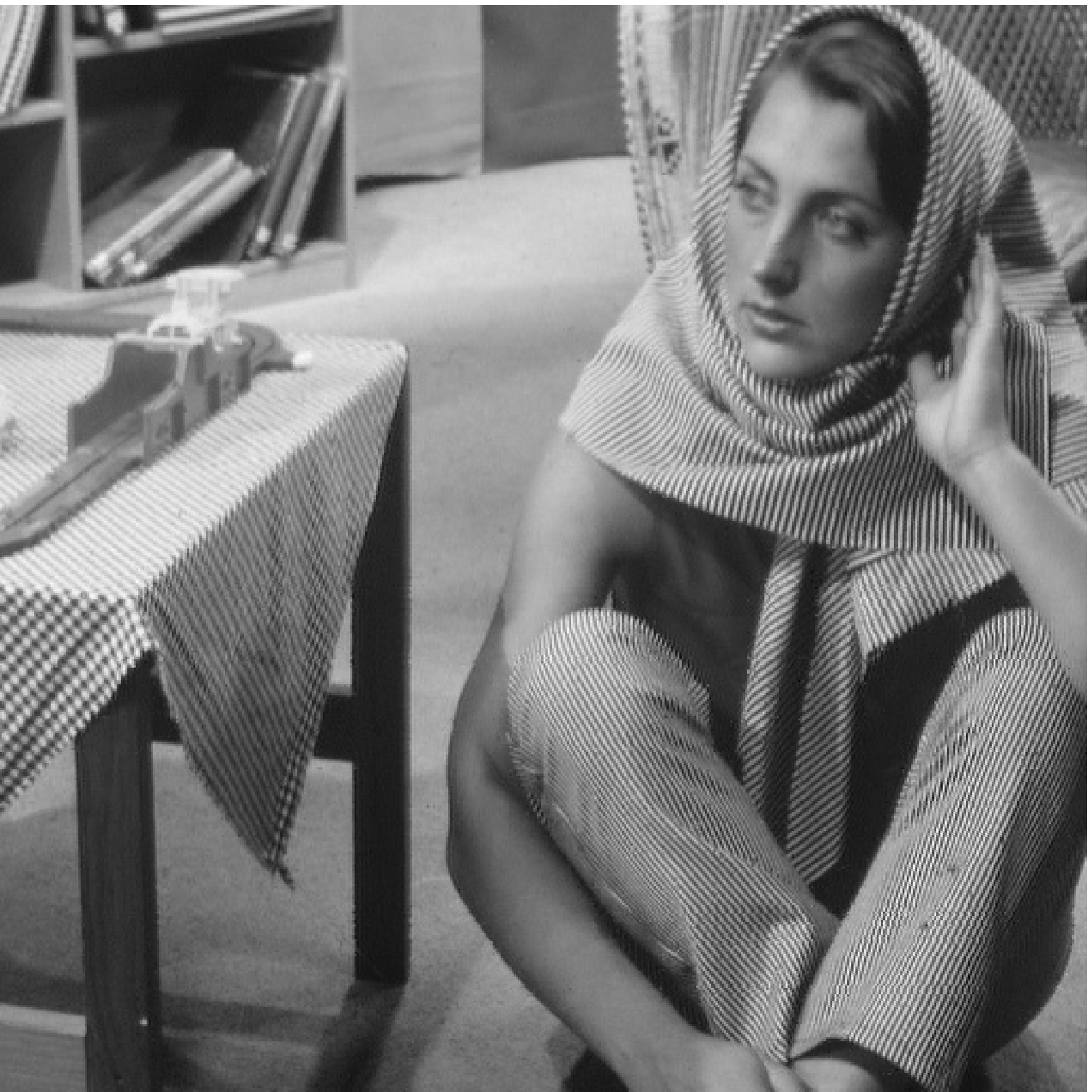}
\includegraphics[width=0.25\textwidth]{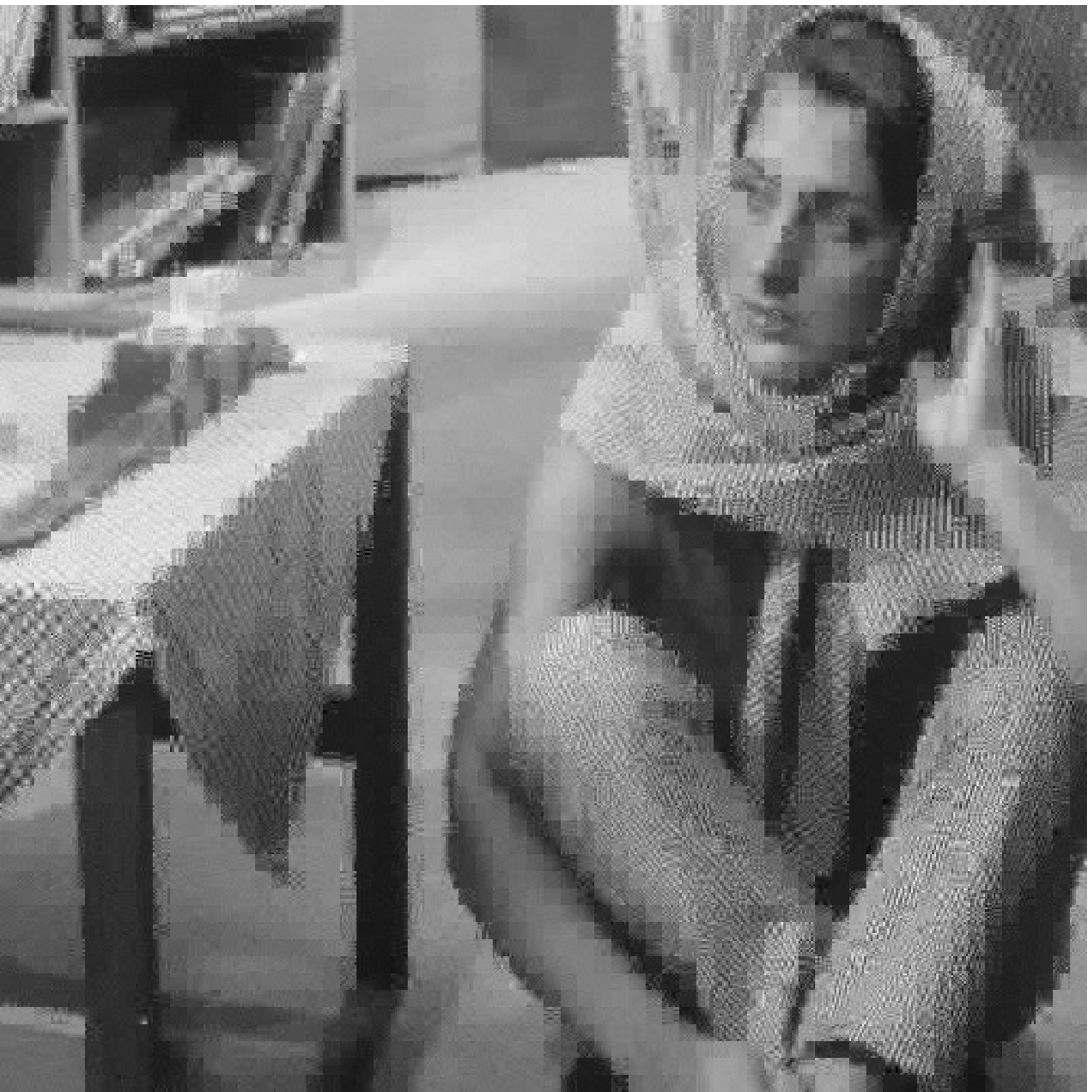}
}\\\medskip 
\mbox{
\includegraphics[width=0.25\textwidth]{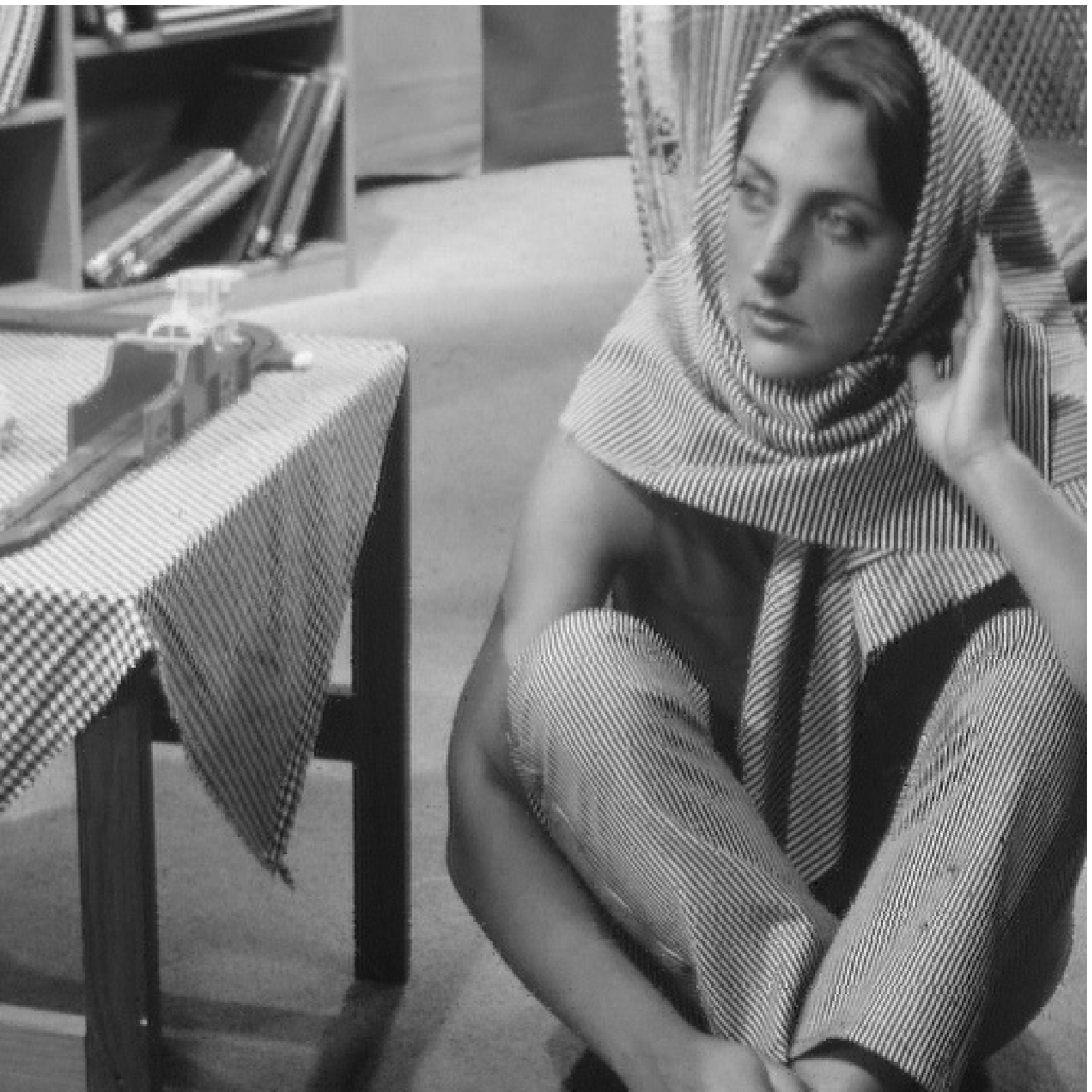}
\includegraphics[width=0.25\textwidth]{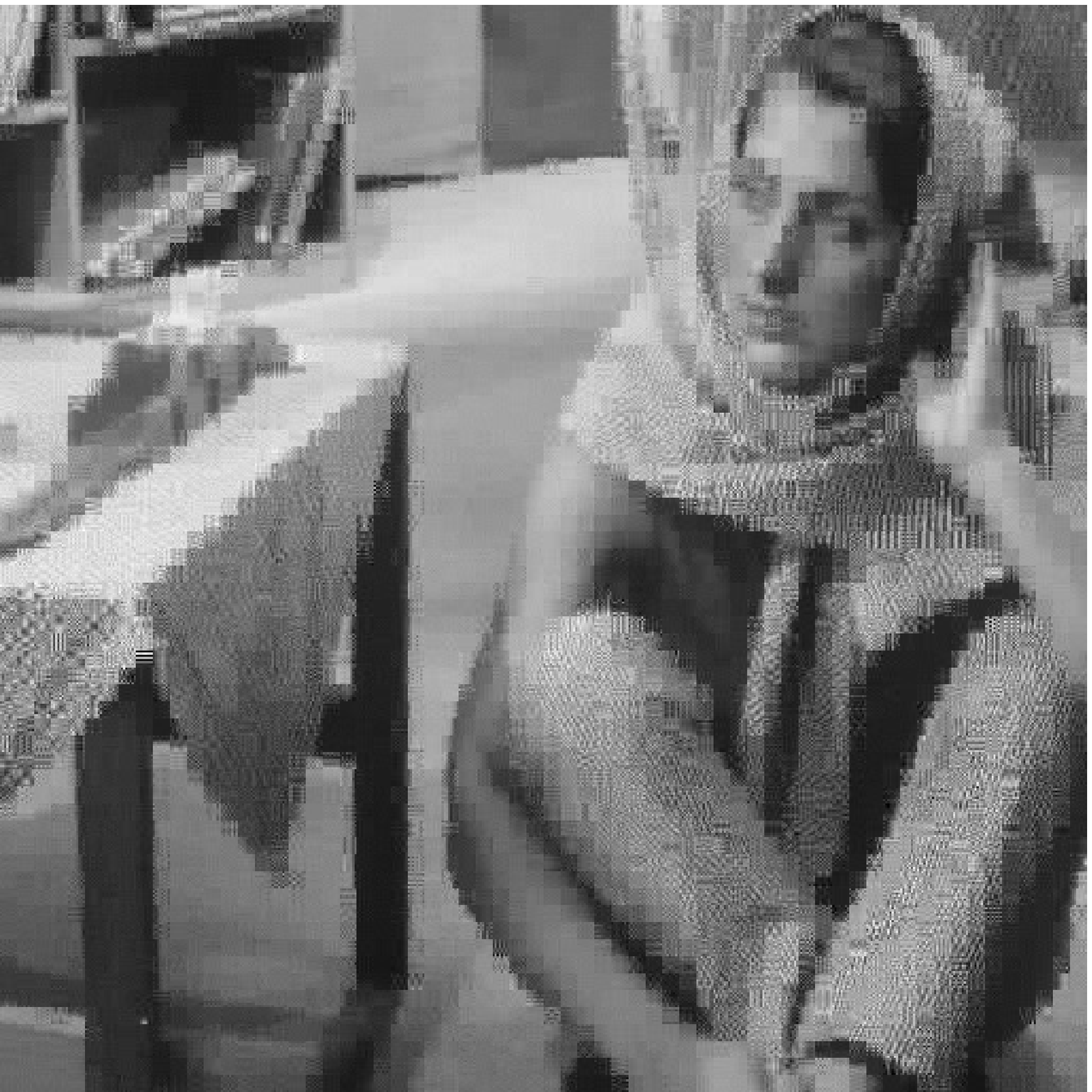}
}
\caption{Filling in pixels missing from Barbara. Top: Original image. Second row from left to right respectively presents images with missing pixels percentage 20\% and 80\% missing pixels.  The images in the third row were derived using the K-SVD dictionary. The images in the fourth row were derived using the learned Parseval dictionary. }
\label{fig:Single_image_miss_pixcel}
\end{center}
\end{figure}

The performance of the dictionaries is presented in Tables \ref{tab:single image miss pixels result} and \ref{tab: miss SSIM}.  We assumed that the mean of an image is known. The mean was subtracted in the experiments and then added to the image to determine the performance.
The images in Figure \ref{fig:Single_image_miss_pixcel} are the results. The perceptual quality of the restored images remains good as long as no more than $60\%$ of the pixels are missing. This is consistent with the PSNR and SSIM performance shown in Tables \ref{tab:single image miss pixels result} and \ref{tab: miss SSIM}. The PSNR and SSIM with missing fractions below $60\%$ exceed $30$ $dB$ and $0.88$, respectively.  The performance of the K-SVD dictionary is uniformly better than that of the Parseval dictionary, as the frame redundancy of the K-SVD dictionary is higher.
\newpage
\begin{table}[!ht]
\begin{center}
\caption{Comparison of average PSNR in experiments involving missing pixels. The values are the average of five experiments using the same percentage of missing pixels. The top row of each block is the PSNR of an image derived from missing pixels. The second row presents the results derived using the K-SVD dictionary, and the third row presents results derived using the learned Parseval dictionary. The PSNR unit is $dB$.} \label{tab:single image miss pixels result}
\begin{tabular}{|c|c||c||c||c||c||c||c|}
\hline
\multicolumn{2}{|c||} {Missing Level} &  Barbara & Man & Lena & Hill & Boat & Average\\\cline{1-8}
\hline\hline     & Corrupted image & 31.17 & 32.87 & 33.68 & 34.00 & 31.95 & 32.73\\                                                
         10\% &K-SVD dictionaries           & \bf{40.37} & \bf{39.69} & \bf{42.85} & \bf{40.48} & \bf{40.20} & \bf{40.72}\\                                                
              & Parseval dictionaries & 38.36 & 38.11 & 40.88 & 39.19 & 38.60 & 39.03\\\cline{1-8}                
\hline\hline     & Corrupted image & 28.16 & 29.81 & 30.68 & 30.96 & 28.99 & 29.72\\
         20\% & K-SVD dictionaries           & \bf{36.02} & \bf{35.89} & \bf{38.95} & \bf{36.73} & \bf{36.28} & \bf{36.78}\\                                                
              & Parseval dictionaries & 34.30 & 34.45 & 37.07 & 35.58 & 34.82 & 35.24\\\cline{1-8}   
\hline\hline     & Corrupted image & 26.38 & 28.08 & 28.92 & 29.20 & 27.25 & 27.97\\
         30\% & K-SVD dictionaries           & \bf{33.07} & \bf{33.42} & \bf{36.36} & \bf{34.32} & \bf{33.62} & \bf{34.16}\\                                                
              & Parseval dictionaries & 31.52 & 32.07 & 34.54 & 33.29 & 32.27	& 32.74\\\cline{1-8}                                
\hline\hline     & Corrupted image & 25.13 & 26.82 & 27.66 & 27.95 & 26.01	& 26.71\\
         40\% & K-SVD dictionaries           & \bf{30.68} & \bf{31.43} &	\bf{34.17} & \bf{32.40} & \bf{31.46} & \bf{32.03}\\                                                
              & Parseval dictionaries & 29.31 & 30.10 & 32.42 & 31.47 & 30.19	& 30.70\\\cline{1-8}
\hline\hline     & Corrupted image & 24.16 & 25.85 &	26.69 & 26.98 & 25.05	& 25.75\\
         50\% & K-SVD dictionaries           & \bf{28.61} & \bf{29.62} &	\bf{32.19} & \bf{30.71} & \bf{29.51}	& \bf{30.13}\\                                                
              & Parseval dictionaries & 27.37 & 28.42 &	30.53 & 29.86 & 28.42	& 28.92\\\cline{1-8}
\hline\hline     & Corrupted image & 23.36 & 25.06 & 25.90 & 26.18 & 24.27	& 24.95\\
         60\% & k-SVD dictionaries           & \bf{26.67} & \bf{27.92} & \bf{30.22} & \bf{29.12} & \bf{27.67}	& \bf{28.32}\\                                                
              & Parseval dictionaries & 25.56 & 26.89 & 28.73 & 28.39 & 26.77	& 27.27\\\cline{1-8}                 
\hline\hline     & Corrupted image & 22.69 & 24.39 & 25.22 & 25.50 & 23.60	& 24.28\\
         70\% & K-SVD dictionaries           & \bf{24.82} & \bf{26.24} & \bf{28.17} & \bf{27.53} & \bf{25.83}	& \bf{26.52}\\                                                
              & Parseval dictionaries & 23.90 & 25.44 & 26.98 & 26.99 & 25.17	& 25.69\\\cline{1-8}                 
\hline\hline     & Corrupted image & 22.11 & 23.80 & 24.64 & 24.92 & 23.02	& 23.70\\
         80\% & K-SVD dictionaries           & \bf{22.98} & \bf{24.51} & \bf{26.02} & \bf{25.79} & \bf{23.94}	& \bf{24.65}\\                                                
              & Parseval dictionaries & 22.36 & 24.05 & 25.30 & 25.51 & 23.59	& 24.16\\\cline{1-8}
\end{tabular}
\end{center}
\end{table}

\newpage
\begin{table}[!ht]
\begin{center}
\caption{Comparison of average SSIM in experiments involving missing pixels. The values are the average of five experiments using the same percentage of missing pixels. The top row of each block is the SSIM of an image derived from missing pixels. The second row presents the results derived using the K-SVD dictionary and the third row presents results derived using the learned Parseval dictionary. } \label{tab: miss SSIM}
\begin{tabular}{|c|c||c||c||c||c||c||c|}
\hline
\multicolumn{2}{|c||} {Missing Level} &  Barbara & Man & Lena & Hill & Boat & Average\\\cline{1-8}
\hline\hline     & Corrupted image & 0.94 & 0.94	& 0.95 & 0.95	 & 0.94	& 0.95\\                                                
         10\% & K-SVD dictionaries           & \bf{0.99} & \bf{0.98}	& \bf{0.99} & \bf{0.98}	 & \bf{0.99}	& \bf{0.99}\\                                                
              & Parseval dictionaries & \bf{0.99} & \bf{0.98}	& 0.98 & \bf{0.98}	 & 0.98	& 0.98\\\cline{1-8}                
\hline\hline     & Corrupted image & 0.89 & 0.89	& 0.91 & 0.89	 & 0.89	& 0.90\\
         20\% & K-SVD dictionaries           & \bf{0.98} & \bf{0.97}	& \bf{0.97} &\bf{ 0.96}	 & \bf{0.96}	&\bf{ 0.97}\\                                                
              & Parseval dictionaries & 0.97 & 0.95	& \bf{0.97} & 0.95	 & \bf{0.96}	& 0.96\\\cline{1-8}   
\hline\hline     & Corrupted image & 0.84 & 0.85	& 0.87 & 0.85	 & 0.85	& 0.85\\
         30\% & K-SVD dictionaries           & \bf{0.96} & \bf{0.94}	& \bf{0.96} & \bf{0.93}	 & \bf{0.94}	& \bf{0.95}\\                                                
              & Parseval dictionarie & 0.94 & 0.92	& 0.94 & 0.92	 & 0.93	& 0.93\\\cline{1-8}                                
\hline\hline     & Corrupted image & 0.80 & 0.80	& 0.83 & 0.80	 & 0.80	& 0.81\\
         40\% & K-SVD dictionaries           & \bf{0.93} & \bf{0.91}	& \bf{0.94} & \bf{0.90}	 & \bf{0.91}	& \bf{0.92}\\                                                
              & Parseval dictionaries & 0.91 & 0.89	& 0.92 & 0.88	 & 0.89	& 0.90\\\cline{1-8}
\hline\hline     & Corrupted image & 0.75 & 0.76	& 0.80 & 0.75	 & 0.75	& 0.76\\
         50\% & K-SVD dictionaries           &\bf{ 0.90} & \bf{0.87}	& \bf{0.91} & \bf{0.86}	 & \bf{0.87}	& \bf{0.88}\\                                                
              & Parseval dictionaries & 0.87 & 0.84	& 0.88 & 0.84	 & 0.84	& 0.85\\\cline{1-8}
\hline\hline     & Corrupted image & 0.70 & 0.71	& 0.77 & 0.70	 & 0.71	& 0.72\\
         60\% & K-SVD dictionaries           & \bf{0.85} & \bf{0.82}	& \bf{0.87} & \bf{0.81}	 & \bf{0.81}	& \bf{0.83}\\                                                
              & Parseval dictionarie & 0.81 & 0.78	& 0.84 & 0.78	 & 0.79	& 0.80\\\cline{1-8}                 
\hline\hline     & Corrupted image & 0.65 & 0.67	& 0.73 & 0.65	 & 0.66	& 0.67\\
         70\% & K-SVD dictionaries           & \bf{0.77} & \bf{0.75}	& \bf{0.82} & \bf{0.74}	 & \bf{0.74}	& \bf{0.76}\\                                                
              & Parseval dictionaries & 0.73 & 0.71	& 0.79 & 0.71	 & 0.72	& 0.73\\\cline{1-8}                 
\hline\hline     & Corrupted image & 0.59 & 0.62	& 0.70 & 0.60	 & 0.61	& 0.62\\
         80\% & K-SVD dictionaries           & \bf{0.67} & \bf{0.65}	& \bf{0.75} & \bf{0.64}	 & \bf{0.64}	& \bf{0.67}\\                                                
              & Parseval dictionaries & 0.63 & 0.63	& 0.72 & 0.63	 & 0.63	& 0.65\\\cline{1-8}
\end{tabular}
\end{center}
\end{table}
\newpage

\newpage
\section{Conclusions} \label{conclusion}

Frames theory provides the foundation for the design of linear operators used in the decomposition and reconstruction of signals.
In this paper, we sought to formulate a dual frame design in which the sparse vector obtained through the decomposition of a signal is also the sparse solution representing the signals that use a reconstruction frame. Our findings demonstrate that such a dual frame does not exist for over-complete frames. Nonetheless, the best approximation to the sparse synthesis solution can be derived from the analysis coefficient using the canonical dual frame.
We took advantage of the analysis and synthesis views of signal representation from the frame perspective and proposed optimization formulations for  problems pertaining to image recovery. We then compared the performance of the solutions derived using dictionaries with different frame bounds. Our results revealed a correlation between recovered images and the frame bounds of dictionaries, thereby demonstrating the importance of using different dictionaries for different applications.

\end{document}